\documentclass[aps,floatfix,preprint,nofootinbib,prd]{revtex4}
\pdfoutput=1
\usepackage[parfill]{parskip}    
\usepackage{graphicx}
\usepackage{amssymb}
\usepackage{epstopdf}
\usepackage{verbatim}
\usepackage{tikz}
\usepackage{slashed}
\usepackage{ulem}
\usepackage{url}
\usepackage[hidelinks]{hyperref}

\newcommand{\be}{\begin{equation}}
\newcommand{\bea}{\begin{eqnarray}}
\newcommand{\beq}[1]{\begin{equation}\label{#1}}
\newcommand{\eq}[1]{Eq. (\ref{#1})}
\newcommand{\eqs}[2]{Eqs. (\ref{#1}) and (\ref{#2})}
\newcommand{\eqss}[3]{Eqs. (\ref{#1}), (\ref{#2}) and (\ref{#3})}
\newcommand{\ee}{\end{equation}}
\newcommand{\eea}{\end{eqnarray}}
\newcommand{\eeq}{\end{equation}}
\newcommand{\lsim}{\!\mathrel{\hbox{\rlap{\lower.55ex \hbox{$\sim$}} \kern-.34em \raise.4ex \hbox{$<$}}}}
\newcommand{\gsim}{\!\mathrel{\hbox{\rlap{\lower.55ex \hbox{$\sim$}} \kern-.34em \raise.4ex \hbox{$>$}}}}

\usepackage{amsmath}
\usepackage{slashed}
\usepackage{float}
\usepackage{mathrsfs}

\newcommand{\Lag}{\mathcal{L}}
\newcommand{\sw}{s_W}
\newcommand{\cw}{c_W}

\newcommand{\se}{s_{\epsilon}}
\newcommand{\ce}{c_{\epsilon}}
\newcommand{\te}{t_{\epsilon}}
\newcommand{\sd}{s_D}
\newcommand{\cd}{c_D}
\newcommand{\td}{t_D}
\newcommand{\mzp}{m_{Z'}}
\newcommand{\mchi}{m_{\text{DM}}}
\newcommand{\mphi}{m_{\text{DM}}}
\newcommand{\mpsi}{m_{\text{DM}}}
\newcommand{\mdm}{m_{\text{DM}}}

\newcommand{\Lf}{\Lag_{\text{DM}}^{\text{(dipole)}}}
\newcommand{\Ls}{\Lag_{\text{DM}}^{\text{(scalar)}}}
\newcommand{\sigmav}{\langle\sigma v\rangle}

\begin{document}
\setlength{\baselineskip}{0.22in}

\begin{flushright}MCTP-14-11 \\
\end{flushright}
\vspace{0.2cm}

\title{Hidden Dipole Dark Matter}
\author{Aaron Pierce and Zhengkang Zhang}
\vspace{0.2cm}
\affiliation{Michigan Center for Theoretical Physics (MCTP) \\
Department of Physics, University of Michigan, Ann Arbor, MI
48109}
\date{\today}

\begin{abstract}
We consider models where a hidden $U(1)'$ interacts with the Standard Model via kinetic mixing. We assume the dark matter is neutral under this
$U(1)'$, but interacts with it via higher dimension operators.  In particular, we consider a hidden dipole operator for fermionic dark matter, and charge radius and Rayleigh operators for scalar dark matter.  These models naturally explain the absence of direct detection signals, but allow for a thermal cosmology.  LHC searches for the $Z'$ represent a powerful probe.
\end{abstract}

\maketitle

\section{Introduction}\label{sec:intro}
The simplest models of weak scale dark matter have direct couplings to the Higgs and electroweak gauge bosons of the Standard Model (SM).  Rapidly improving detection experiments place strong constraints on such models, although windows remain~\cite{Cohen:2011ec,Cheung:2012qy}.  However, it is possible that these models are overly simplistic.  Minor augmentations of a dark sector can allow for thermal relic abundance, but without constraints from direct detection experiments.   Yet these models might be accessible via the Large Hadron Collider (LHC).  Here, we explore one such class of models.  We remain agnostic about the detailed dynamics of the dark sector, but specify a simple portal between the dark sector and the visible sector:  we imagine a new abelian gauge group  $U(1)'$  which kinetically mixes with the SM $U(1)_Y$, possibly induced by heavy particles charged under both groups~\cite{Holdom:1985ag, Dienes:1996zr}.

In the most studied such models, the dark matter is a Dirac fermion with non-zero charge under $U(1)'$, see e.g.~\cite{ArkaniHamed:2008qn,Batell:2009vb,Chun:2010ve,Mambrini:2011dw,Kearney:2013xwa}. The combined constraints from precision electroweak, relic abundance, direct and indirect detection, and collider physics exclude large regions of parameter space.  We will imagine that the dark matter is not, in fact, an elementary charged particle under $U(1)'$. But even so, it may couple to the $U(1)'$ via higher dimensional operators.   This possibility can be studied in an effective field theory (EFT) below some cutoff scale $\Lambda$.  If the dark matter is a Dirac fermion, the leading operator is a dipole interaction:
\begin{equation}
\label{eq:dipole}
    \Lf = i\bar\chi\slashed{\partial}\chi - \mdm\bar\chi\chi + \frac{1}{\Lambda}\bar\chi\sigma^{\mu\nu}\chi\hat Z'_{\mu\nu}.
\end{equation}
And if the dark matter is a complex scalar, the leading operators are the charge radius operator and the Rayleigh operator, provided we impose $\phi$ number conservation\footnote{Another operator $(\phi^2+\phi^{*2})\hat Z'^{\mu\nu}\hat Z'_{\mu\nu}$ can arise if this restriction is lifted.}:
\begin{equation}
\label{eq:scalar}
   \Ls = \partial^{\mu}\phi^*\partial_{\mu}\phi - \mdm^2\phi^*\phi + \frac{1}{\Lambda^2} \left( \kappa_{\text{C}} i\partial^{\mu}\phi^*\partial^{\nu}\phi\hat Z'_{\mu\nu} + \kappa_{\text{R}} \frac{1}{4} \phi^*\phi\hat Z'^{\mu\nu}\hat Z'_{\mu\nu}\right).
\end{equation}
We emphasize that these couplings are to the $U(1)'$ and not to SM gauge bosons. See~\cite{Chang:2010en,Alves:2009nf} for related works that discuss such a possibility, but focus instead on inelastic scattering in direct detection due to a small mass splitting in the dark sector.

Operators as in \eqs{eq:dipole}{eq:scalar} could arise, for example, if the dark matter were a composite particle (see e.g.~\cite{Kribs:2009fy}), neutral under $U(1)'$, but with constituents charged under $U(1)'$.  In this case $\Lambda$ can be interpreted as a compositeness scale.\footnote{\label{footnoteDC}Effective charge operators of significant size may arise (see \eq{eqn:effchg}) based on naive dimensional analysis (NDA)~\cite{Manohar:1983md,Georgi:1992dw}. This could be in tension with direct detection null results; see Section \ref{sec:dd}. So, not just any UV completion will do.  For example, if the dark matter gets its dipole via couplings to particles charged under $U(1)'$ that receive most of their mass from a source other than $U(1)'$ breaking, the effective charge operators may be sufficiently suppressed.} This is analagous to the neutron or the hydrogen atom -- despite their neutrality, they interact with the photon.

In this paper we study these two dark matter candidates. First, we introduce the models and work out the interactions in Section \ref{sec:models}. We proceed to discuss the cosmology, precision electroweak constraints, and collider observables in Sections \ref{relic}-\ref{sec:LHC}. LHC searches for a $Z'$ resonance prove to be particularly powerful. Direct detection is discussed next in Section \ref{sec:dd}. As written, the derivatives present in Eqs.~(\ref{eq:dipole}) and (\ref{eq:scalar}) explain the absence of direct detection signals, both now and into the future.  Thus, these models suffer less tension with constraints than those where the dark matter is charged under $U(1)'$.   However, it is possible that additional higher-dimensional operators (not relevant for the thermal history) can give rise to observable direct detection signals. Throughout this work, our focus is on the window where the dark matter is relatively heavy, say from 100-1000 GeV.  We do, however, comment briefly on the possibility of lighter dark matter for the scalar model in Section \ref{sec:light}. Finally in Section \ref{sec:concl} we conclude.

\section{The Models}\label{sec:models}
We augment the dark sector Lagrangians (either Eq.~(\ref{eq:dipole}) or Eq.~(\ref{eq:scalar})) by allowing kinetic mixing of the $U(1)'$, whose gauge boson is $\hat Z'$ (hat denotes gauge eigenstate field), with the SM hypercharge boson $\hat B$:
    \begin{equation}
    \Lag = \Lag_{\text{SM}} + \frac{\se}{2}\hat B_{\mu\nu}\hat Z'^{\mu\nu} - \frac{1}{4}\hat Z'_{\mu\nu}\hat Z'^{\mu\nu} + \frac{1}{2}m_{\hat Z'}^2\hat Z'_{\mu}\hat Z'^{\mu} + \Lag_{\text{DM}}.
    \end{equation}
Here $\se$, short for $\sin\epsilon$, parameterizes the kinetic mixing. The mass term for $\hat Z'$ arises from the vacuum expectation value (vev) of a dark Higgs field. We assume the dark Higgs boson does not significantly impact the phenomenology.  This would be the case, for example, if the dark Higgs boson were heavy with respect to the other dark sector particles.

The mass terms can be diagonalized (and kinetic terms made canonical) by a rotation from the $(\hat Z, \hat A, \hat Z')$ basis to the mass eigenstate basis $(Z, A, Z')$:
    \begin{eqnarray}\label{rotation}
    \hat Z^{\mu} &=& (\cd+\sd\te\sw)Z^{\mu} + (\sd-\cd\te\sw)Z'^{\mu},\\
    \hat A^{\mu} &=& A^{\mu} - \sd\te\cw Z^{\mu} + \cd\te\cw Z'^{\mu}, \label{rotation2}\\
    \hat Z'^{\mu} &=& \frac{\cd}{\ce} Z'^{\mu} - \frac{\sd}{\ce} Z^{\mu}\label{rotation3},
    \end{eqnarray}
    where $\ce\equiv\cos\epsilon$, $\cd\equiv\cos\theta_D$, etc.
The angle $\theta_D$ is given by
    \begin{equation}
    \tan2\theta_D = \frac{\sw\sin2\epsilon}{\ce^2-\se^2\sw^2-\frac{m_{\hat Z'}^2}{m_{\hat Z}^2}}.
    \end{equation}
    After eliminating $m_{\hat Z}$, $m_{\hat Z'}$ in favor of the masses of the mass eigenstate fields $m_Z$, $m_{Z'}$ through
    \begin{eqnarray}
    m_{\hat Z}^2 &=& \frac{m_Z^2}{1+\td\te\sw},\\
    m_{\hat Z'}^2 &=& m_{Z'}^2\ce^2(1+\td\te\sw),
    \end{eqnarray}
    we find
    \begin{equation}\label{tdr}
    \td = \frac{1-r \pm \sqrt{(1-r)^2-4\te^2\sw^2 r}}{2\te\sw r},
    \end{equation}
    where $r=\frac{m_{Z'}^2}{m_Z^2}$, and $+$ ($-$) is taken for $r\geq r_+(\epsilon)$ $\bigl(r\leq r_-(\epsilon)\bigr)$, with
    \begin{equation}
    r_{\pm}(\epsilon) = 1 + 2\te^2\sw^2 \pm 2\sqrt{\te^2\sw^2(1+\te^2\sw^2)}.
    \end{equation}
    $r_-(\epsilon) < r < r_+(\epsilon)$ is not possible.
In the limit $\epsilon\ll1$ and $\mzp\gg m_Z$,
\begin{equation}
\theta_{D} \simeq -\epsilon \sw\frac{m_Z^2}{\mzp^2}.
\end{equation}

    Due to the mixing among the neutral gauge bosons shown in Eqs. (\ref{rotation})-(\ref{rotation3}), the $Z'$ acquires $\mathcal{O}(\epsilon)$ couplings to the SM neutral currents:
    \beq{currents}
    \begin{split}
    \Lag &\supset - \hat Z_{\mu}J_Z^{\mu} - \hat A_{\mu}J_{\text{EM}}^{\mu}\\
    &= - Z_{\mu}\Bigl[(\cd+\sd\te\sw)J_Z^{\mu}-\sd\te\cw J_{\text{EM}}^{\mu}\Bigr] - A_{\mu}J_{\text{EM}}^{\mu} - Z'_{\mu}\Bigl[(\sd-\cd\te\sw)J_Z^{\mu}+\cd\te\cw J_{\text{EM}}^{\mu}\Bigr]\\
    &\simeq  - Z_{\mu} J_Z^{\mu} - A_{\mu}J_{\text{EM}}^{\mu} +\epsilon Z'_{\mu}\Bigl[\sw\Bigl(1+\frac{m_Z^2}{\mzp^2}\Bigr) J_Z^{\mu} - \cw J_{\text{EM}}^{\mu}\Bigr].
   \end{split}
    \eeq
The last expression holds in the limit of small $\epsilon$ and large $\mzp$.

After the rotation into the mass eigenstate basis, the interactions of Eq.~(\ref{eq:dipole}) or Eq.~(\ref{eq:scalar}) induce couplings of the dark matter to the gauge bosons:

    \begin{eqnarray}
    \text{Dipole model:}\qquad \Lag_{\chi\chi Z'} &=& \frac{1}{\Lambda}\left(\frac{\cd}{\ce}\right)\bar\chi\sigma^{\mu\nu}\chi Z'_{\mu\nu},\\
    \Lag_{\chi\chi Z} &=& -\frac{1}{\Lambda}\left(\frac{\sd}{\ce}\right)\bar\chi\sigma^{\mu\nu}\chi Z_{\mu\nu};\\
    \text{Scalar model:}\qquad \Lag_{\phi\phi Z'} &=& \frac{\kappa_{\text{C}}}{\Lambda^2}\left(\frac{\cd}{\ce}\right)i\partial^{\mu}\phi^*\partial^{\nu}\phi Z'_{\mu\nu},\\
    \Lag_{\phi\phi Z} &=& -\frac{\kappa_{\text{C}}}{\Lambda^2}\left(\frac{\sd}{\ce}\right)i\partial^{\mu}\phi^*\partial^{\nu}\phi Z_{\mu\nu}\\
    \Lag_{\phi\phi Z'Z'} &=& \frac{\kappa_{\text{R}}}{4\Lambda^2}\left(\frac{\cd}{\ce}\right)^2\phi^*\phi Z'^{\mu\nu} Z'_{\mu\nu},\\
    \Lag_{\phi\phi Z'Z} &=& -\frac{\kappa_{\text{R}}}{2\Lambda^2}\left(\frac{\cd}{\ce}\right)\left(\frac{\sd}{\ce}\right)\phi^*\phi Z'^{\mu\nu} Z_{\mu\nu,}\\
    \Lag_{\phi\phi ZZ} &=& \frac{\kappa_{\text{R}}}{4\Lambda^2}\left(\frac{\sd}{\ce}\right)^2\phi^*\phi Z^{\mu\nu} Z_{\mu\nu}.
    \end{eqnarray}
Notably, there is no dark matter-photon coupling.

    \section{Relic Abundance of Dark Matter}\label{relic}
\begin{figure}[t]
    \centering
    \makebox[6in]{\underline{{\normalsize Dipole model}}}\\[5pt]
    \includegraphics[width=3.2in]{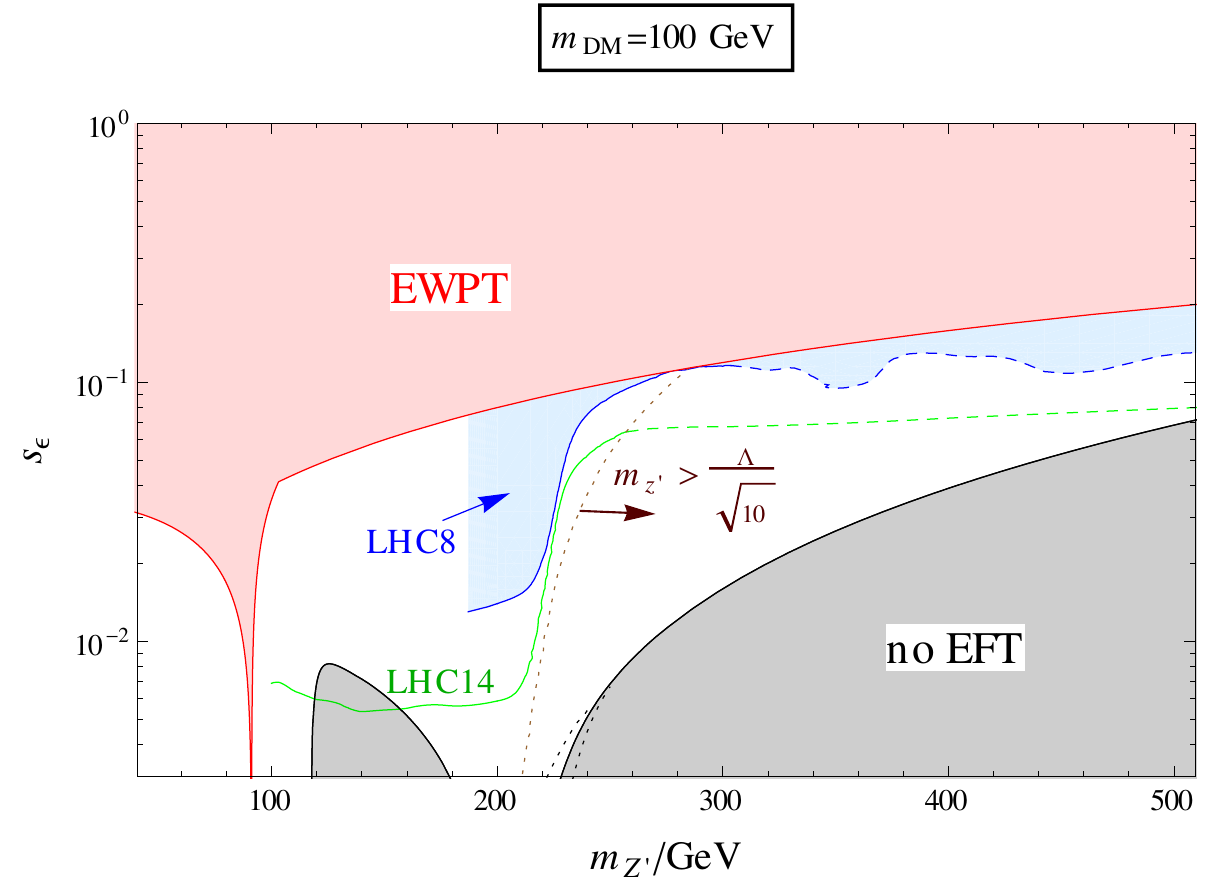}
    \includegraphics[width=3.2in]{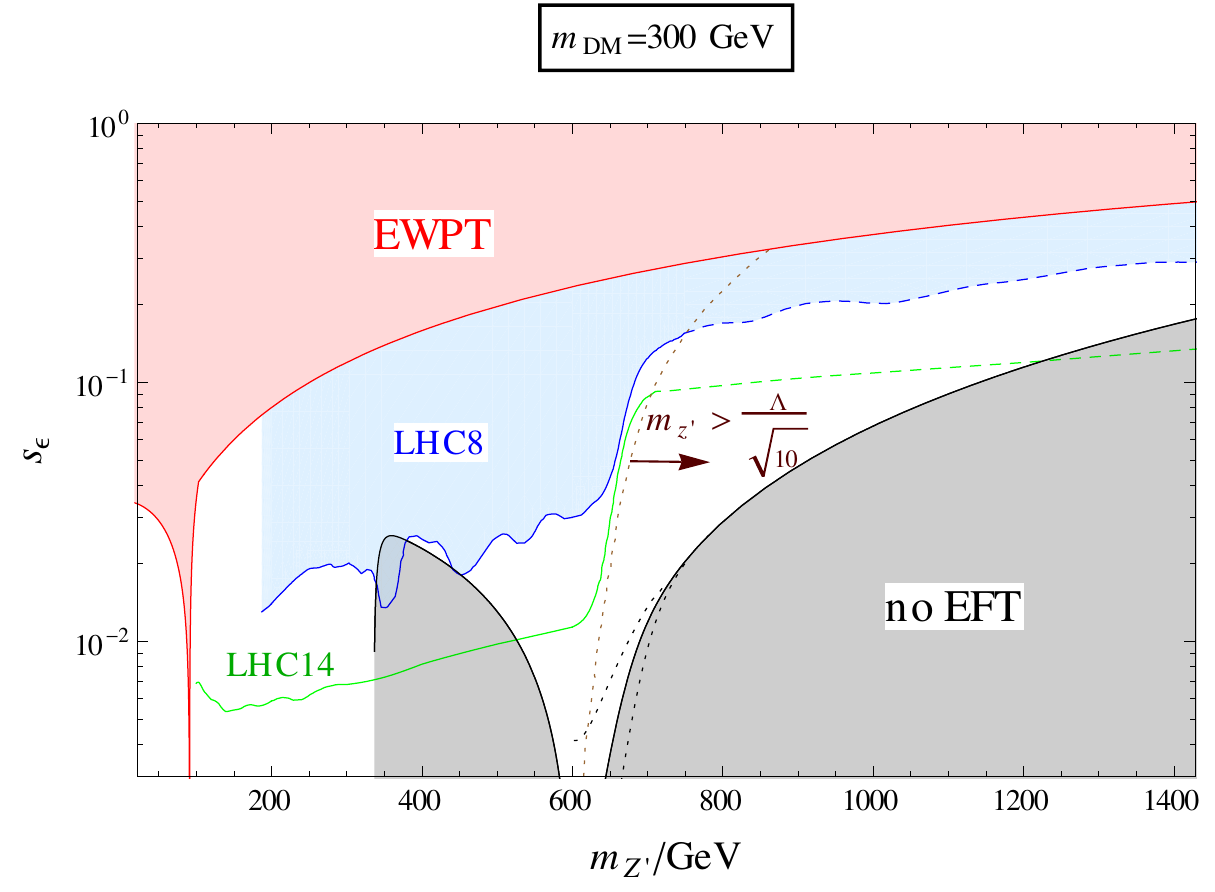}\\[3pt]
    \includegraphics[width=3.2in]{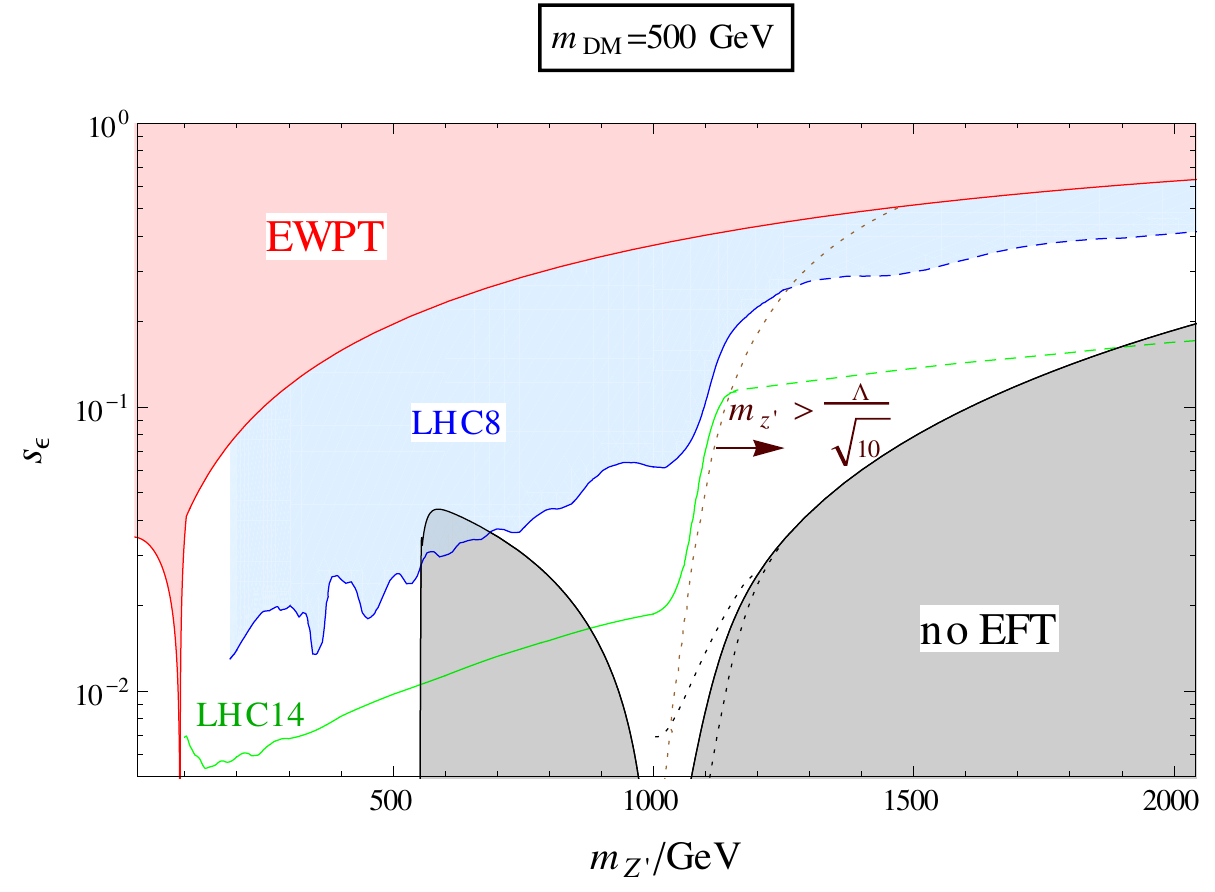}
    \includegraphics[width=3.2in]{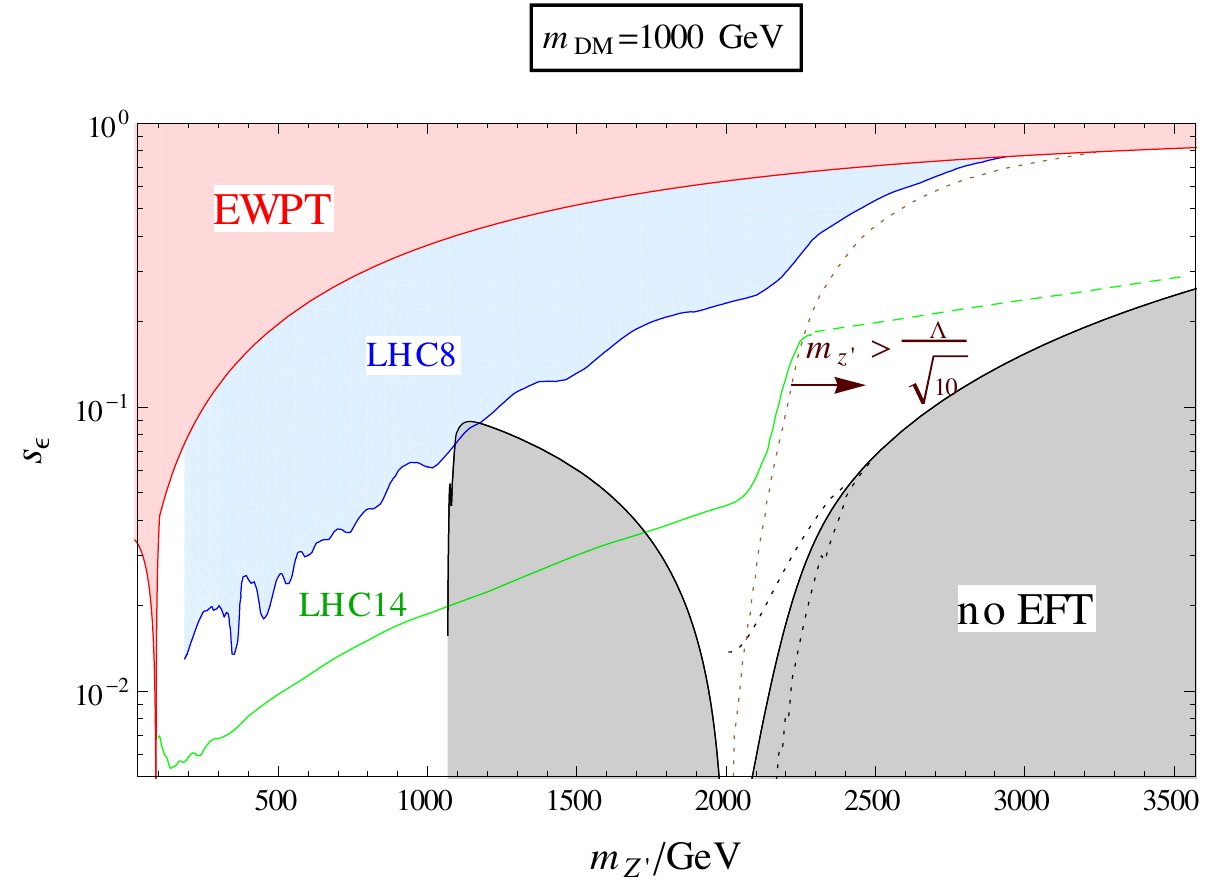}
\caption{Allowed parameter space for the dipole model in the $(\mzp,\se)$ plane for four fixed values of $\mchi$. Regions where the EFT breaks down for the $\Lambda$ required to give the observed relic abundance are shaded, as are those excluded by EWPT and LHC searches for the $Z'$ at 8 TeV. Also shown are projected exclusion limits at the 14 TeV LHC (300 fb$^{-1}$ for a single experiment). Uncertainties are associated with the boundary of the ``no EFT'' regions near $\mzp\sim2\mdm$ due to unknown $\Gamma_{Z'\to\text{Dark}}$; the solid curves are obtained assuming $\frac{\Gamma_{Z'\to\text{Dark}}}{\mzp}=10^{-2}$, while dotted curves correspond to $\frac{\Gamma_{Z'\to\text{Dark}}}{\mzp}=10^{-1}$ (upper), $10^{-3}$ (lower). Also, the exact LHC limits are called into question for $\mzp>\frac{\Lambda}{\sqrt{10}}$, where $\Gamma_{Z'\to\text{Dark}}$ is not calculable in the EFT. The limits shown in dashed curves are obtained following the prescriptions explained in the text.}
    \label{dipoleplots}
    \end{figure}

    There are four free parameters in the dipole model: $\se$, $\mzp$, $\mdm$, and $\Lambda$. Consistency with the observed thermal relic abundance of dark matter $\Omega h^2=0.1199\pm0.0027$~\cite{Ade:2013zuv} provides one constraint among the four parameters. After the relic abundance is fixed to the central value, three free parameters remain, which we take to be $\se$, $\mzp$ and $\mdm$. If $\mdm$ is held fixed, $\Lambda$ becomes a function on the $(\mzp,\se)$ plane. To describe annihilations consistently within the EFT, we require $\Lambda>\mdm$.  In Fig.~\ref{dipoleplots} we shade the regions in the $(\mzp,\se)$ plane where $\Lambda$, as determined by the relic abundance, is smaller than $\mdm$ (labeled ``no EFT''). We perform calculations of the relic abundance using micrOMEGAs~\cite{Belanger:2013oya}.

  \begin{figure}[t]
    \centering
    \makebox[6in]{\underline{{\normalsize Scalar model}}}\\[5pt]
    \includegraphics[width=3.2in]{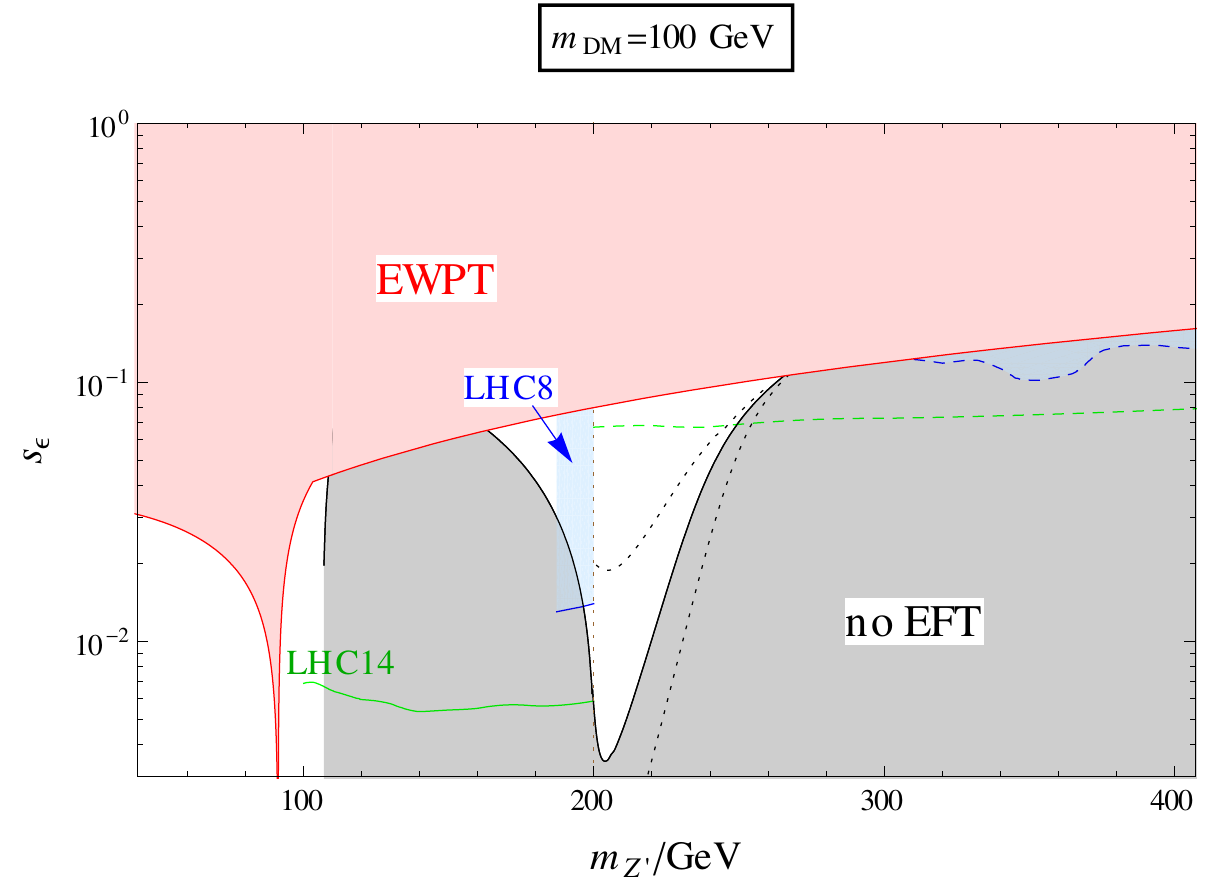}
    \includegraphics[width=3.2in]{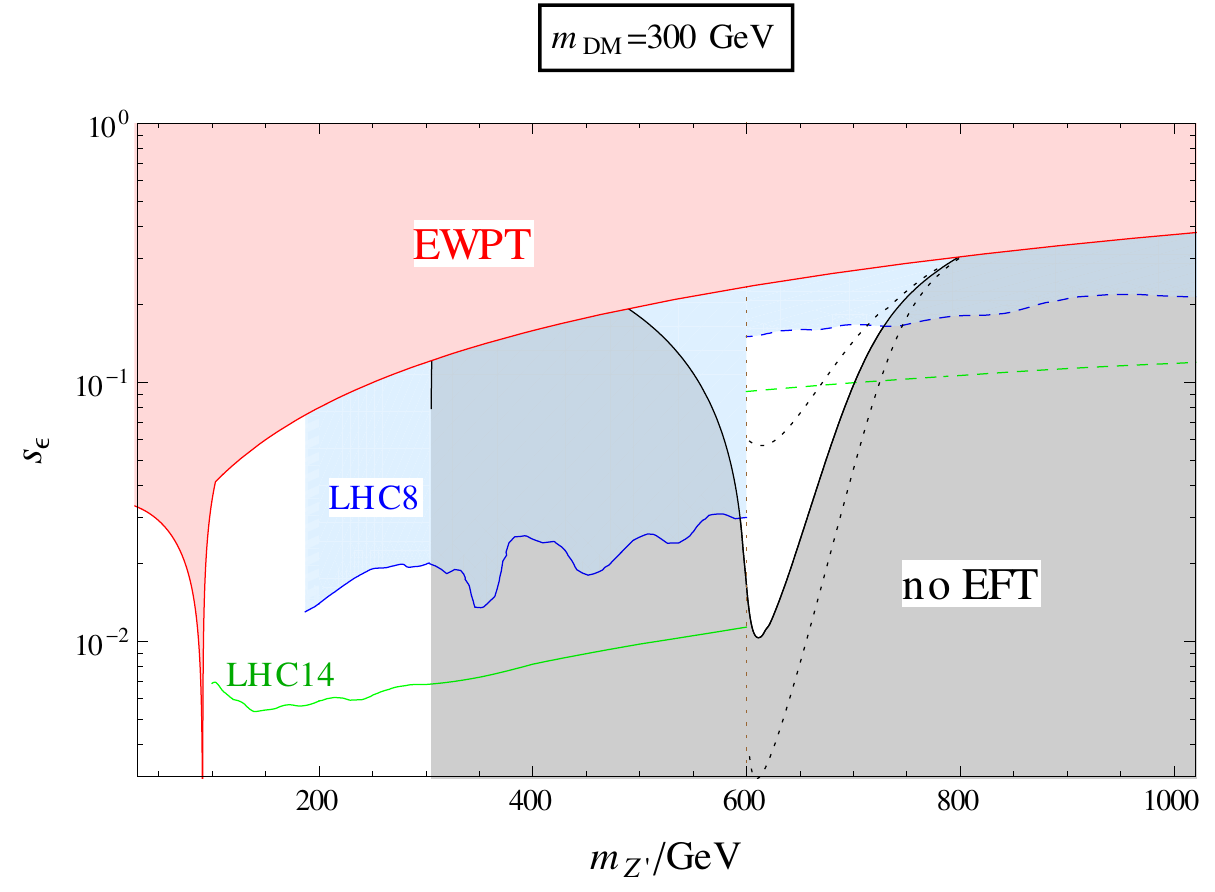}\\[3pt]
    \includegraphics[width=3.2in]{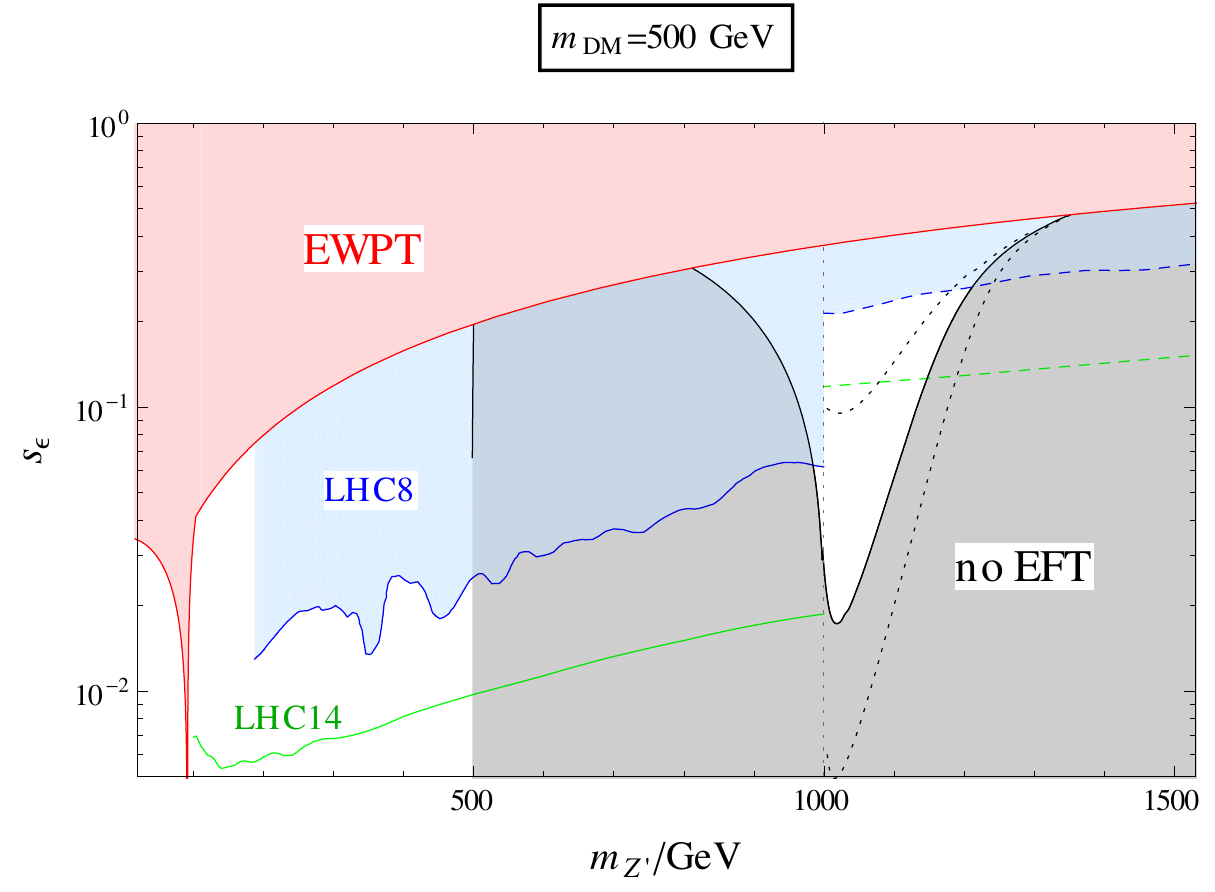}
    \includegraphics[width=3.2in]{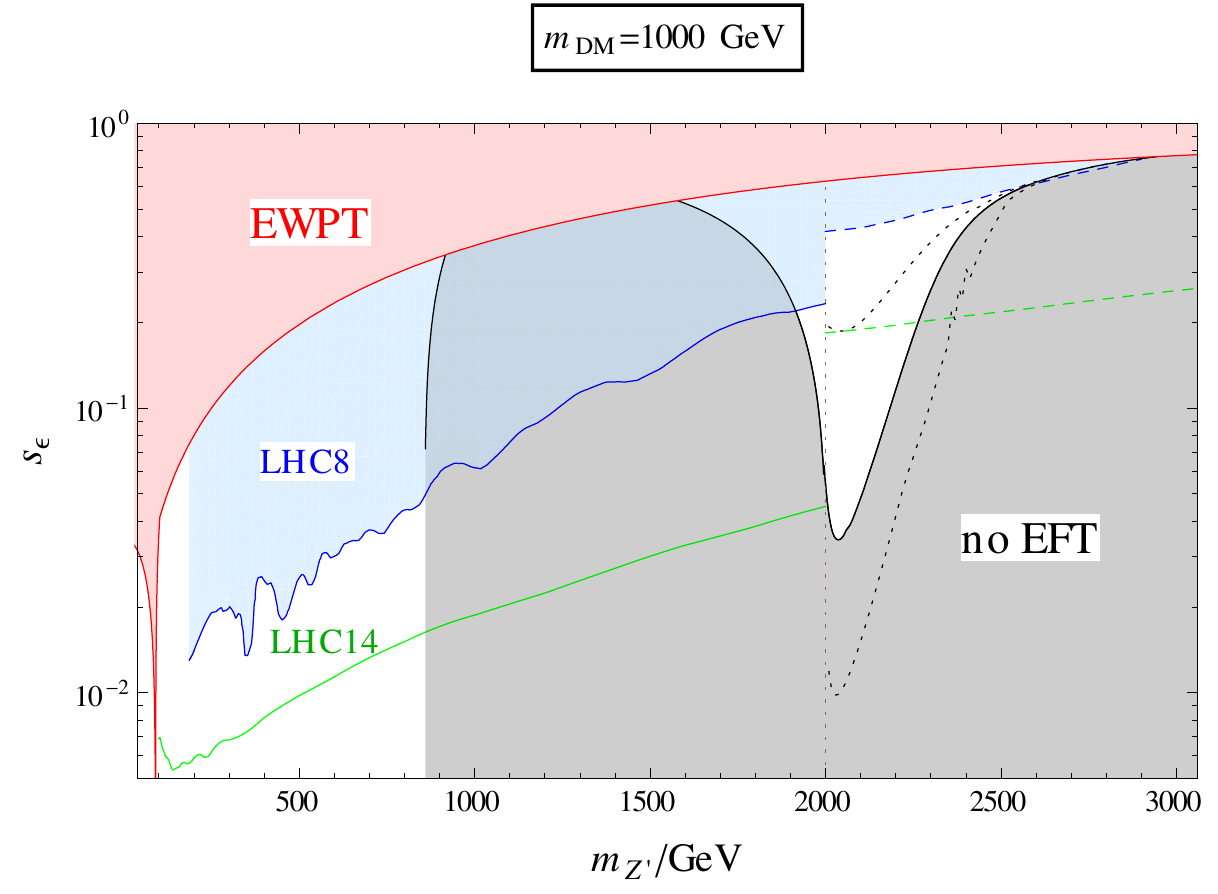}
\caption{Allowed parameter space for the scalar model in the $(\mzp,\se)$ plane for four fixed values of $\mdm$, assuming $\kappa_{\text{C}}=\kappa_{\text{R}}=1$. Regions where the EFT breaks down for the $\Lambda$ required to give the observed relic abundance are shaded, as are those excluded by EWPT and LHC searches for the $Z'$ at 8 TeV. Also shown are projected exclusion limits at the 14 TeV LHC (300 fb$^{-1}$ for a single experiment). Uncertainties are associated with the boundary of the ``no EFT'' regions near $\mzp\sim2\mdm$ due to unknown $\Gamma_{Z'\to\text{Dark}}$; the solid curves are obtained assuming $\frac{\Gamma_{Z'\to\text{Dark}}}{\mzp}=10^{-2}$, while dotted curves correspond to $\frac{\Gamma_{Z'\to\text{Dark}}}{\mzp}=10^{-1}$ (upper), $10^{-3}$ (lower). Also, the exact LHC limits are called into question for $\mzp>\frac{\Lambda}{\sqrt{10}}$ (essentially for $\mzp>2\mdm$), where $\Gamma_{Z'\to\text{Dark}}$ is not calculable in the EFT. The limits shown in dashed curves are obtained following the prescription $\frac{\Gamma_{Z'\to\text{Dark}}}{\mzp}=10^{-2}$.}
    \label{chgrdsplots}
    \end{figure}

For sufficiently small $\se$, we see $\Lambda$ typically becomes smaller than $\mdm$, and the validity of the EFT is called into question.  This tension arises because the annihilation channel $\chi\bar\chi\to f\bar f$, which dominates over much of the parameter space, has a cross section proportional to $\frac{\epsilon^2}{\Lambda^2}$ for $\epsilon$ small. For $\mzp\sim2\mdm$ the cross section is greatly enhanced by resonance, and smaller $\se$  may be accommodated. Also,  for $\mzp<\mchi$, the annihilation channel $\chi\bar\chi\to Z'Z'$, not subject to $\epsilon$ suppression, is open. Thus, in this regime,  small $\se$ is consistent with cosmology.

A similar story holds for the scalar model; see Fig. \ref{chgrdsplots}. Here, a slight complication arises due to an additional parameter $\kappa_{\text{R}}/\kappa_{\text{C}}$. In Fig. \ref{chgrdsplots} we set $\kappa_{\text{R}}/\kappa_{\text{C}}=1$, but it is straightforward to extrapolate to a wide range of $\kappa_{\text{R}}/\kappa_{\text{C}}$ because the charge radius and Rayleigh operators contribute differently to the cosmology.
\begin{itemize}
  \item For $\mzp<\mphi$, the Rayleigh operator dominates. This can be seen from the thermally averaged cross section times velocity $\sigmav$ of the dominant annihilation channel $\phi\bar\phi\to Z'Z'$:
      \be
      \begin{split}
      \sigmav_{\phi\bar\phi\to Z'Z'} &\simeq \sigmav_{\phi\bar\phi\to Z'Z'}^{s\text{-wave}}\\ &=
      \begin{cases}
        \frac{1}{64\pi\mphi^2} \left(\frac{\cd}{\ce}\right)^4 \left(\frac{\mphi}{\Lambda}\right)^8 \frac{x^4(1-x^2)^{5/2}}{(1-x^2/2)^2} &\text{(charge radius)},\\
        \frac{1}{8\pi\mphi^2} \left(\frac{\cd}{\ce}\right)^4 \left(\frac{\mphi}{\Lambda}\right)^4 (1-x^2+3x^4/8)(1-x^2)^{1/2} &\text{(Rayleigh)}, \label{rayleighZZ}
      \end{cases}
      \end{split}
      \ee
      where $x\equiv\frac{\mzp}{\mphi}$, and the first (second) line of the last equation is obtained for $\kappa_{\text{C}}=1, \kappa_{\text{R}}=0$ ($\kappa_{\text{C}}=0, \kappa_{\text{R}}=1$). The smaller prefactor and the suppression as $x\to0$ or $x\to1$ typically render the charge radius operator subdominant to the Rayleigh operator in this regime. Actually, much of the regime $\mzp<\mphi$ would be inaccessible within the range of validity of the EFT if only the charge radius operator were present, for it under-annihilates the dark matter. We see in Fig. \ref{chgrdsplots} that the presence of the Rayleigh operator largely lifts any constraints in this regime. The exception occurs for the largest $\mphi$ considered, where $\sigmav\propto\mphi^{-2}$ is suppressed, requiring smaller $\Lambda$ to compensate -- this excludes the region $\mzp\gtrsim850$ GeV in the $\mphi=1000\text{ GeV}$ plot. We note no similar excluded region exists in the dipole model even for $\mdm = 1000$ GeV as
      \be \label{dipoleZZ}
\begin{split}
       \sigmav_{\chi\bar\chi\to Z'Z'}&\simeq \sigmav_{\chi\bar\chi\to Z'Z'}^{s\text{-wave}} \\
        &=\frac{2}{\pi\mchi^2} \left(\frac{\cd}{\ce}\right)^4 \left(\frac{\mchi}{\Lambda}\right)^4 \frac{(1-x^2)^{3/2}(1+2x^2+5x^4/16)}{(1-x^2/2)^2} \,\,\,\,\text{(dipole)}
\end{split}
      \ee
is intrinsically larger.

To get a feel for the size of the higher dimensional operators required in this regime, we can take the $x \to 0$  limit of Eqs.~(\ref{rayleighZZ}) and (\ref{dipoleZZ}), and invert them for $\Lambda$. Assuming $\sigmav^{s\text{-wave}}$=2.3$\times10^{-26}\text{ cm}^3/\text{s}$, we find
\be \label{dipolesize}
  \frac{\mdm}{\Lambda} \simeq
      \begin{cases}
7.5 \times 10^{-2} \left( \frac{\mdm}{100 \text{\, GeV}}\right)^{1/2}        &\text{(dipole),}\\
1.5 \times 10^{-1} \left( \frac{\mdm}{100 \text{\, GeV}}\right)^{1/2}       &\text{(Rayleigh).}
      \end{cases}
\ee
These values suggest that new states charged under $U(1)'$ lie near the weak scale.
  \item For $\mzp>\mphi$, where $\phi\bar\phi\to Z'Z'$ is kinematically forbidden, the charge radius operator dominates. This is true even though $\phi\bar\phi\to f\bar f$ induced by the charge radius operator is $p$-wave suppressed, because the large multiplicity of $f\bar f$ final states allows this channel to dominate over the $s$-wave $\phi\bar\phi\to ZZ$ and $\phi\bar\phi\to ZZ'$, the only channels induced by the Rayleigh operator. The $p$-wave suppression is due to the mismatch of angular momentum between the $s$-wave initial state $\phi\bar\phi$ ($J=0$) and the $s$-channel vector boson ($J=1$).  Compensating for the $p$-wave suppression requires smaller  $\Lambda$, and thus implies a smaller region in the $(\mzp,\se)$ plane is accessible within the EFT than for the dipole model, which has an $s$-wave piece.
\end{itemize}

As a final comment, the width of $Z'$ is relevant for the calculation of the relic abundance in the resonance region as it regulates the $Z'$ propagator. But for $\mzp>2\mdm$, $Z'$ may decay into the dark sector with partial width $\Gamma_{Z'\to\text{Dark}}$. A naive calculation using $\Lf$ (or $\Ls$) may not give the right result for $\Gamma_{Z'\to\text{Dark}}$, because if $\mdm<\Lambda\lesssim\mzp$, states in the hidden sector with masses above $\Lambda$ may be relevant, or $\chi$ (or $\phi$) may not even be the right degrees of freedom for the calculation. Indeed, we can calculate $\Gamma_{Z'\to\text{Dark}}$ in the EFT without worrying about the UV completion only if $\mzp\ll\Lambda$. This is certainly not the case on the $\Lambda=\mdm$ contours, where $\mzp>2\Lambda$, so $\Gamma_{Z'\to\text{Dark}}$ has to be prescribed. The solid curves in Figs.~\ref{dipoleplots} and \ref{chgrdsplots} are obtained assuming $\frac{\Gamma_{Z'\to\text{Dark}}}{\mzp}=10^{-2}$. Two other choices are shown in these plots -- the dotted curves correspond to $\frac{\Gamma_{Z'\to\text{Dark}}}{\mzp}=10^{-1},10^{-3}$. We see that the regions excluded by $\Lambda<\mdm$ are not significantly affected by different prescriptions for the dipole model. For the scalar model, the impact of $\Gamma_{Z'\to\text{Dark}}$ is larger due to the limited size of surviving parameter space near $\mzp\sim2\mdm$. We also remark that other parts of the relic abundance calculation are still valid for $\mzp\gtrsim\Lambda$, since the energy scale of interest in the thermal freeze-out process is $\mdm<\Lambda$.

\section{Precision Electroweak Constraints}\label{sec:ewpt}

    Electroweak precision tests (EWPT) put upper limits on $\se$ as a function of $\mzp$. This has been studied in~\cite{Kumar:2006gm,Hook:2010tw,Chun:2010ve}. For simplicity, we adopt the analytical result of~\cite{Kumar:2006gm} for a heavy $Z'$:
    \begin{equation}
    \left(\frac{\te}{0.1}\right)^2\left(\frac{250\text{GeV}}{\mzp}\right)^2 \lesssim 1.
    \end{equation}
    This bound is conservative, but not much weaker than that obtained in~\cite{Hook:2010tw} through a more complete analysis. The $\rho$ parameter alone provides a strong constraint near the Z pole, for which we use the result in~\cite{Chun:2010ve}. Patching the two together by taking the envelope of these regions\footnote{This simplistic approach gives a reasonable approximation to the full result in~\cite{Hook:2010tw}.},  EWPT exclude the red shaded regions in the $(\mzp,\se)$ plane in Figs.~\ref{dipoleplots} and \ref{chgrdsplots}.

    \section{LHC limits and future probes}\label{sec:LHC}
A $Z'$ with couplings to SM fermions can show up at colliders as a resonance in the dilepton invariant mass spectrum. The non-observation of a resonance at the LHC leads to an upper limit on the $Z'$ production cross section~\cite{CMS:2013qca,ATLAS:2013jma}, which in turn constrains the parameters of our models. We choose to focus on the dimuon channel because, at present, the CMS dimuon channel provides the strongest constraint.

    In the narrow width approximation (NWA) , the cross section factorizes into a product of the $Z'$ resonance production cross section and the branching ratio. This factorization holds exactly to NLO, and the resonance production cross section has a simple quadratic scaling with the vector and axial couplings of the $Z'$ to the quarks~\cite{Beringer:1900zz,Accomando:2010fz}:
    \begin{equation}\label{pp2Zp2mumu}
    \begin{split}
    \sigma(pp\to Z'X\to\mu^+\mu^-X) &\simeq 
    \sum_q (v_q^2+a_q^2)W_q(s,\mzp^2) BR(Z'\to\mu^+\mu^-)\\
    &\simeq \big[(v_u^2+a_u^2)W_u(s,\mzp^2)+(v_d^2+a_d^2)W_d(s,\mzp^2)\big]BR(Z'\to\mu^+\mu^-).
    \end{split}
    \end{equation}
    Here information from the parton distribution functions (pdf)
is contained in $W_q$, and only $W_u$ and $W_d$ are kept since they are substantially larger than the $W_q$ functions for the other quarks \cite{Beringer:1900zz}. We follow the PYTHIA coupling conventions~\cite{Ciobanu:2005pv}:
    \begin{equation}
    \Lag_{Z'qq} = \frac{e}{4\sw\cw}\bar q \gamma^{\mu}(v_q-a_q\gamma^5) q Z'_{\mu}.
    \end{equation}
    Comparing this with Eq.~(\ref{currents}), we obtain:
    \begin{eqnarray}\label{vqaq}
    a_d &=&  -a_u = \sd - \cd\te\sw,\\
    v_q &=& a_q + 4Q_q (\sd\sw^2-\cd\te\sw).\label{vqaq2}
    \end{eqnarray}
    These are functions of $\se$ and $\mzp$. In the limit of small $\epsilon$ and large $\mzp$,
    \bea
    v_u &\simeq& -\epsilon\sw\left[\tfrac{5}{3} - \left(1-\tfrac{8}{3}\sw^2\right)\tfrac{m_Z^2}{\mzp^2}\right],\\
    a_u &\simeq& \epsilon\sw\left(1+\tfrac{m_Z^2}{\mzp^2}\right),\\
    v_d &\simeq& \epsilon\sw\left[\tfrac{1}{3} - \left(1-\tfrac{4}{3}\sw^2\right)\tfrac{m_Z^2}{\mzp^2}\right],\\
    a_d &\simeq& -\epsilon\sw\left(1+\tfrac{m_Z^2}{\mzp^2}\right).
    \eea

Now with \eq{pp2Zp2mumu}, we can calculate the $Z^{\prime}$ production cross section in the $(\mzp,\se)$ plane by first calculating it for some reference values of the couplings, and then scaling the result according to Eqs.~(\ref{pp2Zp2mumu}), (\ref{vqaq}), and (\ref{vqaq2}). This is done with PYTHIA 8~\cite{Sjostrand:2006za,Sjostrand:2007gs} using pdf set CTEQ6L1. In addition, the branching ratios are calculated with micrOMEGAs. Comparing the calculated cross sections with the current 95\% CL exclusion limits from CMS~\cite{CMS:2013qca} and ATLAS~\cite{ATLAS:2013jma} experiments at $8\text{ TeV}$ LHC with integrated luminosity $20.6\text{ fb}^{-1}$ and $20\text{ fb}^{-1}$, respectively, we shade the excluded regions above the blue curves in Figs.~\ref{dipoleplots} and \ref{chgrdsplots}.\footnote{CMS has stronger limits than ATLAS, so we use the CMS results for the most part ($300\text{ GeV}<\mzp<3500\text{ GeV}$). The exception is the mass range $190\text{ GeV}\lesssim\mzp<300\text{ GeV}$, where only ATLAS presents limits. The transition between the two at 300 GeV is smooth because their limits roughly equal at this mass.}

There is a potential subtlety when $\mzp>2\mdm$ that requires discussion. In this regime, the branching ratio to muons may be suppresed due to the opening of dark sector channels. We write
    \begin{equation}
    BR(Z'\to\mu^+\mu^-) = \frac{\Gamma_{Z'\to\mu^+\mu^-}}{\Gamma_{Z'\to\text{SM}}+\Gamma_{Z'\to\text{Dark}}} = \frac{BR_0}{1+\frac{\Gamma_{Z'\to\text{Dark}}}{\Gamma_{Z'\to\text{SM}}}},
    \end{equation}
    where $BR_{0}$ represents the branching ratio to muons in the limit $\Gamma_{Z'\to\text{Dark}}\to0$. As discussed at the end of Section \ref{relic}, our models  $\Lf$ and $\Ls$, interpreted as EFTs below the cutoff scale $\Lambda$, may not give a reliable calculation of  $\Gamma_{Z'\to\text{Dark}}$ (and hence $BR$) if $\mzp\ll\Lambda$ is not satisfied. Here we adopt the following rule for the calculation of $BR$: if $\mzp<\frac{\Lambda}{\sqrt{10}}$, the EFT calculation of $\Gamma_{Z'\to\text{Dark}}$ is trusted  (the $\sqrt{10}$ is a somewhat arbitrary numerical factor ensuring that we are comfortably within the regime where the EFT is valid); otherwise, we impose a prescription for $\Gamma_{Z'\to\text{Dark}}$ as outlined below.
    The contours $\mzp=\frac{\Lambda}{\sqrt{10}}$ (with $\Lambda$ determined by relic abundance) are shown as brown dotted curves 
    in Figs.~\ref{dipoleplots} and \ref{chgrdsplots}; call these curves $\mzp=\widetilde m_{Z'}(\se)$. To the right of these curves, the exclusion limits are prescription dependent, and are shown as dashed curves using
prescriptions explained in the following two paragraphs.
Only the solid parts of the exclusion limits (corresponding to calculable $BR$) are to be taken quantitatively.

    For the dipole model, we use the prescription
$\frac{\Gamma_{Z' \to \rm{Dark}}}{m_{Z'}} =\frac{\Gamma_{Z'\to\chi\bar\chi}(\widetilde m_{Z'}(\se),\se)}{\widetilde m_{Z'}(\se)}$ for $\mzp>\widetilde m_{Z'}(\se)$ in Fig.~\ref{dipoleplots}. In other words, we naively extrapolate the value of $\frac{\Gamma_{Z'\to\text{Dark}}}{\mzp}$ on the brown dotted curves (where it is presumably calculable in the EFT) to the right.

  For the scalar model, the extrapolation prescription will not work, because the contours $\mzp=\frac{\Lambda}{\sqrt{10}}$ (where one might trust the EFT calculation) are now indistinguishable from $\mzp=2\mphi$. This results from the $p$-wave suppression, which forces $\Lambda$ to be lower in order to avoid under-annihilation in the early universe. That is, there is no region wherein one trusts the EFT calculation of $\Gamma_{Z'\to\text{Dark}}$ from which we can then extrapolate. Thus, we adopt the alternate (somewhat arbitrary) prescription $\frac{\Gamma_{Z'\to\text{Dark}}}{\mzp}=10^{-2}$ in Fig.~\ref{chgrdsplots}.

  \begin{figure}[t]
    \centering
    \includegraphics[width=6in]{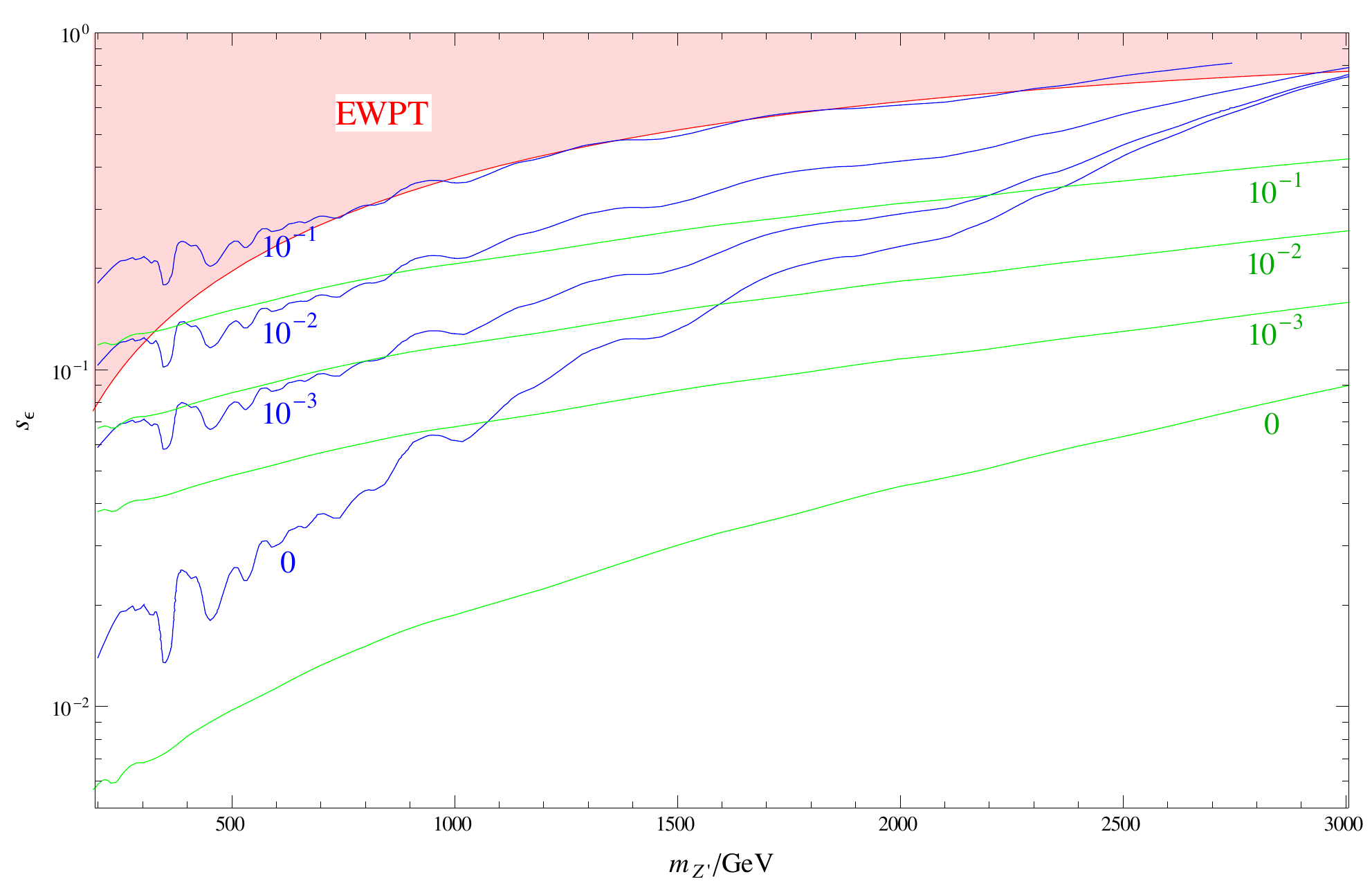}
\caption{LHC exclusion limits obtained assuming the numbers labeled for the value of $\frac{\Gamma_{Z'\to\text{Dark}}}{\mzp}$. Both current (blue, wiggly, labeled on the left) and projected (green, labeled on the right) limits are shown. The actual exclusion curves for a particular $\mdm$ follow the $\frac{\Gamma_{Z'\to\text{Dark}}}{\mzp}=0$ contours for $\mzp<2\mdm$, and deviate from them for $\mzp>2\mdm$.}
    \label{widthplots}
    \end{figure}

To illustrate the sensitivity of the LHC exclusion limits to the $\Gamma_{Z'\to\text{Dark}}$ prescription, we show in Fig.~\ref{widthplots} the exclusion limits obtained assuming $\frac{\Gamma_{Z'\to\text{Dark}}}{\mzp}=0$, $10^{-3}$, $10^{-2}$,  $10^{-1}$, respectively, over a wide range of $\mzp$. For a given $\mdm$, the actual exclusion curve would follow the $\frac{\Gamma_{Z'\to\text{Dark}}}{\mzp}=0$ contour for $\mzp<2\mdm$, where the $Z'$ cannot decay to dark matter, and deviate from it for $\mzp>2\mdm$, as the constraint is weakened by a nonzero $\Gamma_{Z'\to\text{Dark}}$.

    Also shown in Figs.~\ref{dipoleplots}, \ref{chgrdsplots} and \ref{widthplots} are the projected $95\%$ CL exclusion limits (green curves) in future 14 TeV LHC experiments, with $300\text{ fb}^{-1}$ integrated luminosity for the CMS detector.\footnote{One may also be interested in the projected $5\sigma$ discovery limits. These can be obtained by rescaling the projected exclusion curves. For $\mzp<2\mdm$, $S\sim\epsilon^2$. For $\mzp>2\mdm$, the signal $S\sim\epsilon^4$ as long as $Z'\to$ Dark channels dominate (normally this is the case, since $\frac{\Gamma_{Z'\to\text{SM}}}{\mzp}\sim 10^{-4}$ for $\epsilon\sim10^{-1}$).} These are obtained by simulating signal and background events in MadGraph/MadEvent/Pythia/PGS~\cite{Alwall:2011uj,Sjostrand:2006za,Conway:pgs} with pdf set CTEQ6L1, and analyzing the event samples using ExRootAnalysis~\cite{ERA}.
Only the dominant Drell-Yan background is considered. We generate matched samples of events with up to 2 jet emission~\cite{Alwall:2007fs,Alwall:2008qv}, and require $S=\max\{1.96\sqrt{B+(\delta B)^2},3\}$ for the exclusion limits. Here $S$ and $B$ are the number of signal and background events satisfying the CMS event selection criteria stated in~\cite{CMS:2013qca}, with the dimuon invariant mass falling into a bin around the resonant mass $\mzp\pm\Delta(\mzp)$. The bin size is chosen according to the dimuon mass resolution of the CMS detector, which is found in~\cite{Chatrchyan:2012oaa} to be
    \beq{resolution}
    \text{Resolution}= \left[0.01+0.04\left(\frac{\mzp}{\text{TeV}}\right)\right] \mzp \equiv R(\mzp).
    \eeq
     To be explicit, we set $\Delta(\mzp)=\alpha R(\mzp)$, and choose $\alpha$ such that the  strongest limits can be obtained. We find that the optimal $\alpha$ is around 1 for $\mzp\gtrsim 800\text{ GeV}$, and is around 0.5 for $\mzp\lesssim 800\text{ GeV}$.\footnote{Presumably, in the lower mass region, \eq{resolution} overestimates the resolution. It is stated in~\cite{Chatrchyan:2012oaa} that the dimuon mass resolution is 5\% (9\%) at 1 TeV (2 TeV), and increases linearly with dimuon mass. It is not clear this linear extrapolation is valid down to arbitrarily low mass.} The limits are found to be insensitive to the choice of $\alpha$ as long as it is near the optimum. The $\delta B$ term characterizes the uncertainty in the number of background events in the resonance peak.  For concreteness,  $\delta$ is chosen to be 2\%. With this choice, this term becomes relevant for $\mzp\lesssim400\text{ GeV}$. If $\delta=$ 1\% (3\%), this term is relevant for  $\mzp\lesssim$ 250 (600) GeV.  In Figs. \ref{dipoleplots} and \ref{chgrdsplots} we use the same prescriptions in the regime $\mzp>2\mdm$ as for the current exclusion limits described above. In Fig. \ref{widthplots} we show the limits for $\frac{\Gamma_{Z'\to\text{Dark}}}{\mzp}=0$, $10^{-3}$, $10^{-2}$,  $10^{-1}$.

    From Figs.~\ref{dipoleplots} and \ref{chgrdsplots}, we see that for both the dipole model and the scalar model, the upper part of the region $\mzp<2\mchi$ surviving the EWPT and relic abundance constraints has been excluded by LHC experiments, while future experiments can push the bounds even lower. 
    The present published limits constrain a $Z'$ as light as 190 GeV, but our results on the projected limits suggest that even lighter $Z'$ may also be probed.\footnote{Projected limits are shown for $\mzp>100$ GeV. Whether an even lighter $Z'$ can be probed will depend on the $p_{T}$ threshold for the muon trigger.} For $\mzp>2\mchi$ where the $Z'$ may decay predominantly to dark matter, the limits become weaker. But depending on $\Gamma_{Z'\to\text{Dark}}$, a significant part of this region may be probable in the future.

    Three comments are in order before we close this section. First, our calculations are done in the NWA, in accord with the limit set by CMS in~\cite{CMS:2013qca}.\footnote{It is claimed in~\cite{CMS:2013qca} that the limits can be applied to resonances which are not narrow, but since the limit setting procedure in that paper assumes a Breit-Wigner shape whose width is taken to be that of the $Z'_{\psi}$ (which is narrow), it is not obvious that the same limits can be applied to a broad resonance.} Whether this is a good approximation depends largely on the UV completion of our models, which gives $\Gamma_{Z'\to\text{Dark}}$. Second, final state radiation (FSR) can shift the Breit-Wigner shape of the resonance in the dimuon invariant mass spectrum toward lower mass, which is not accounted for in the CMS limits in~\cite{CMS:2013qca}. This may lead to a small error of a few percent for the current exclusion limits (blue) we obtain. FSR is taken into account in our calculation of the projected limits (green). Finally, one might think that our dark matter models may also be probed in the jets + missing $E_T$ channel, especially when the $Z'$ decays invisibly, see e.g.~\cite{Zhou:2013raa}. However, after quantitatively investigating this question, we find that even with 300 fb$^{-1}$ data at the 14 TeV LHC, the constraint from this channel is weak -- even weaker than EWPT.

    \section{Direct detection of dark matter}\label{sec:dd}

In this section we will see that a dark matter particle described solely by $\Lf$ or $\Ls$ is invisible in current and future direct detection experiments. Nevertheless, direct detection may still be relevant for our models if higher dimensional operators with reasonable coefficients are generated from the UV completion. We present an overview of the results in this section, leaving the calculational details to the Appendix.

We find it useful to compare our models with those where the dark matter is directly charged under $U(1)'$. As a reference model, we have
\begin{equation}\label{refmodel}
\Lag_{\text{DM}}^{\text{(reference)}} = i\bar\psi\slashed{\partial}\psi - \mpsi\bar\psi\psi - c'\bar\psi\gamma^{\mu}\psi\hat Z'_{\mu}.
\end{equation}

\begin{figure}[t]
    \centering
    \includegraphics[width=3.2in]{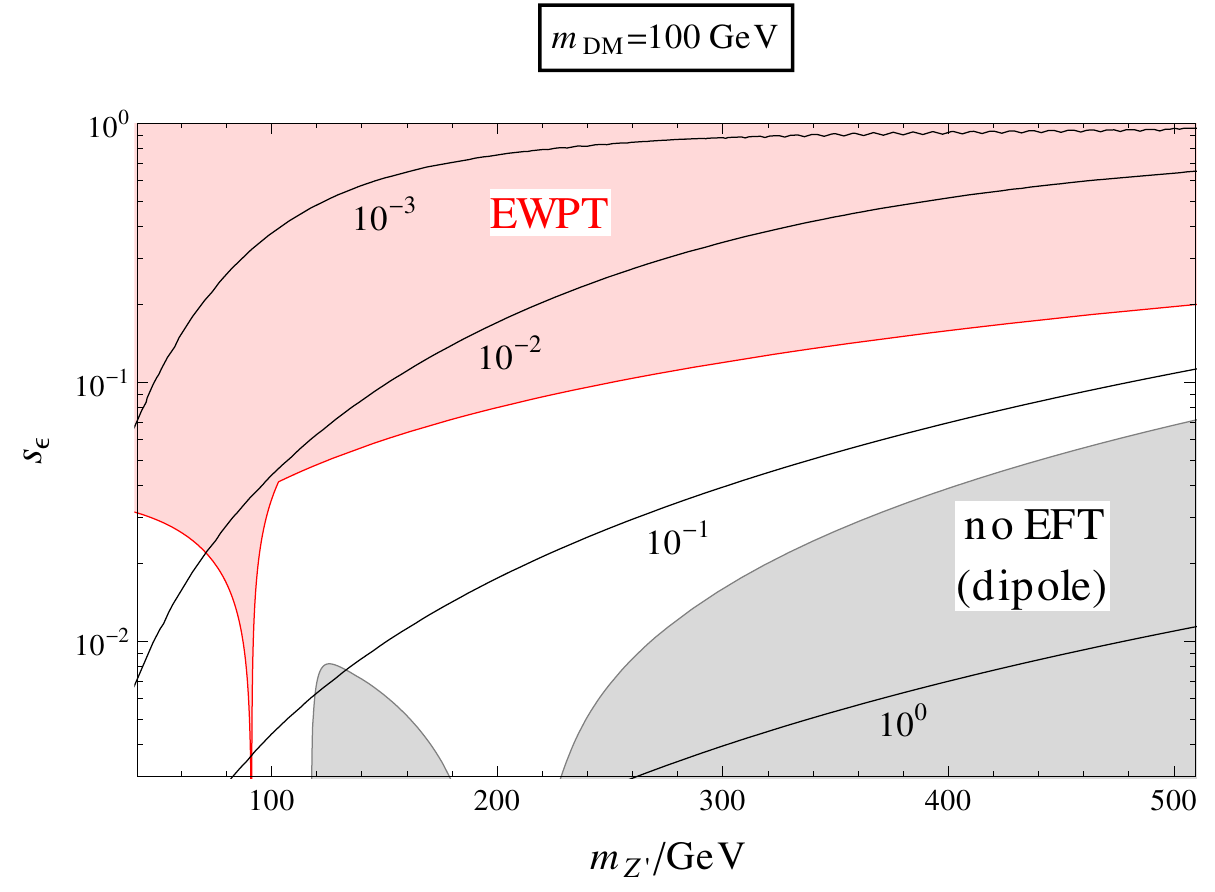}
    \includegraphics[width=3.2in]{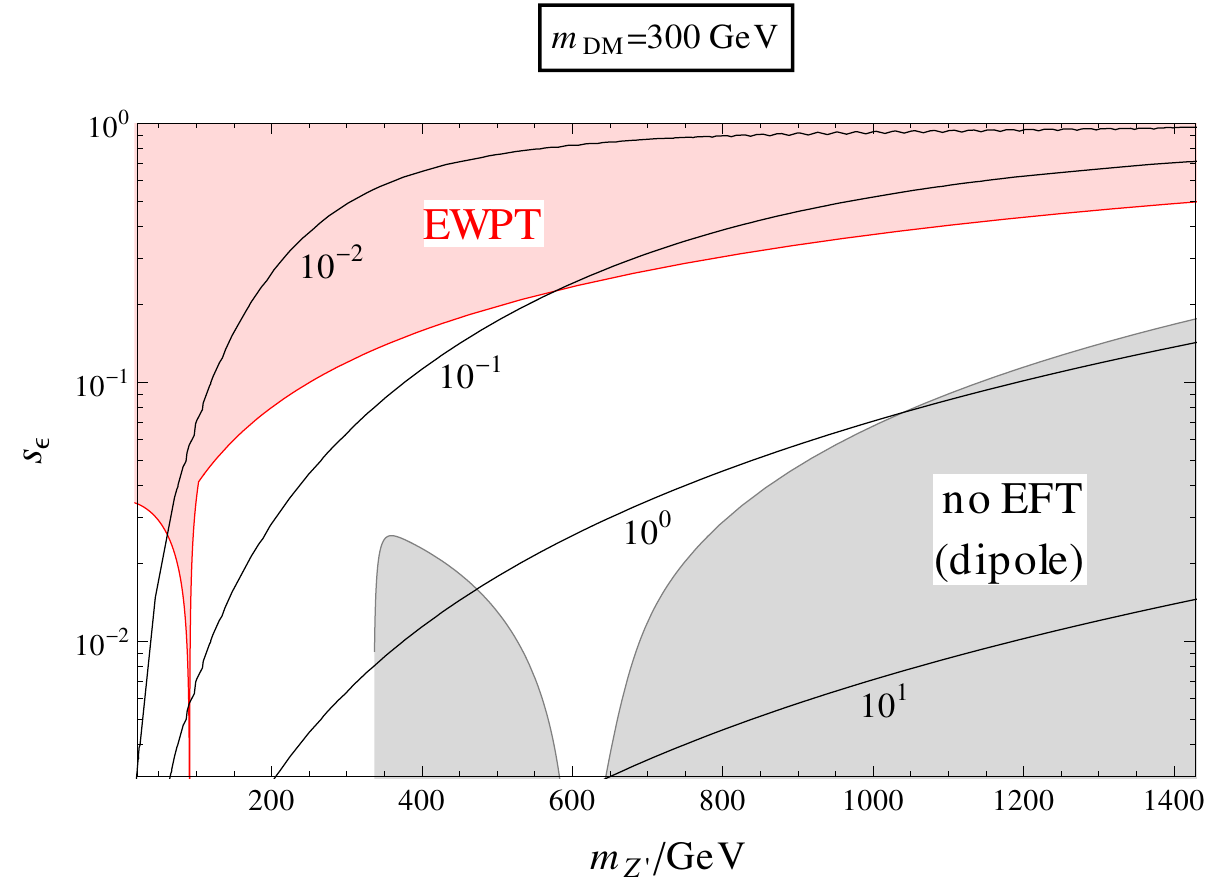}\\[3pt]
    \includegraphics[width=3.2in]{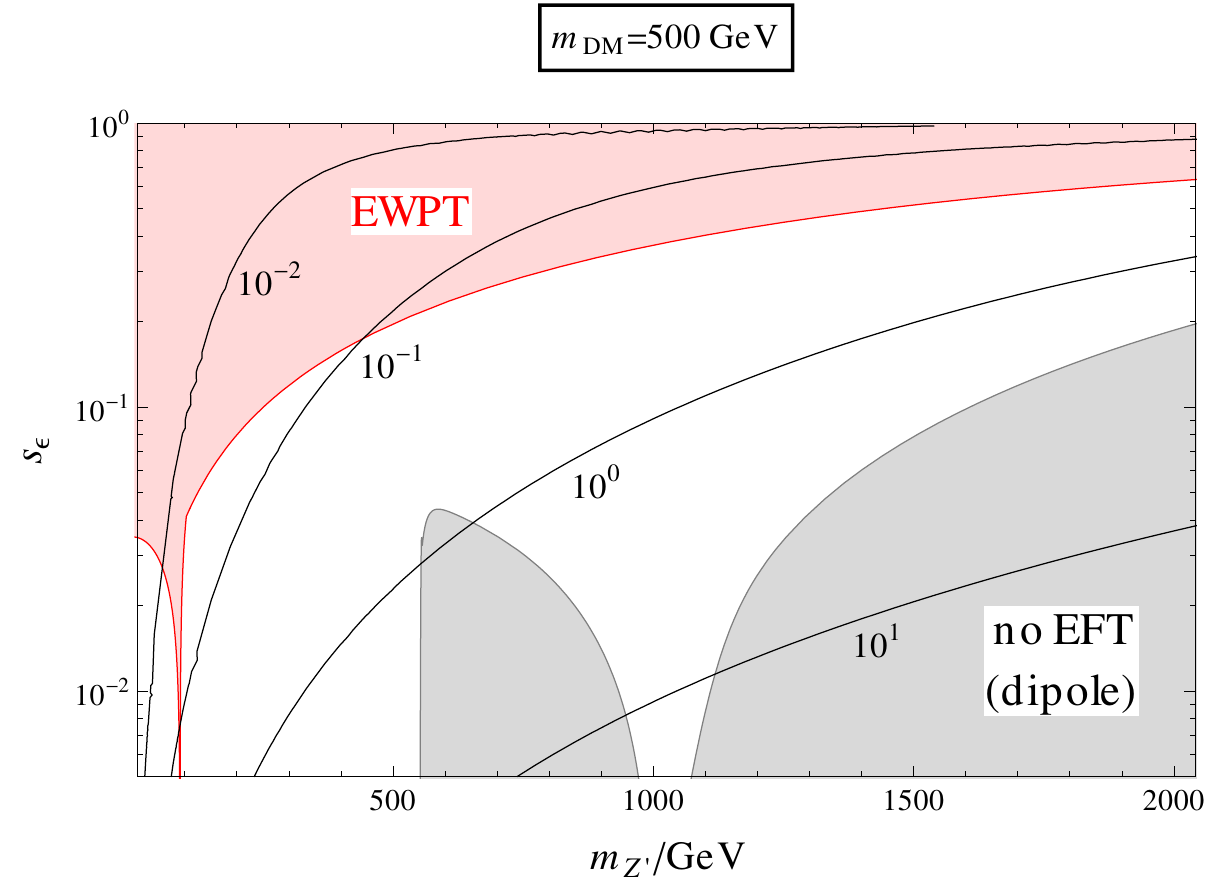}
    \includegraphics[width=3.2in]{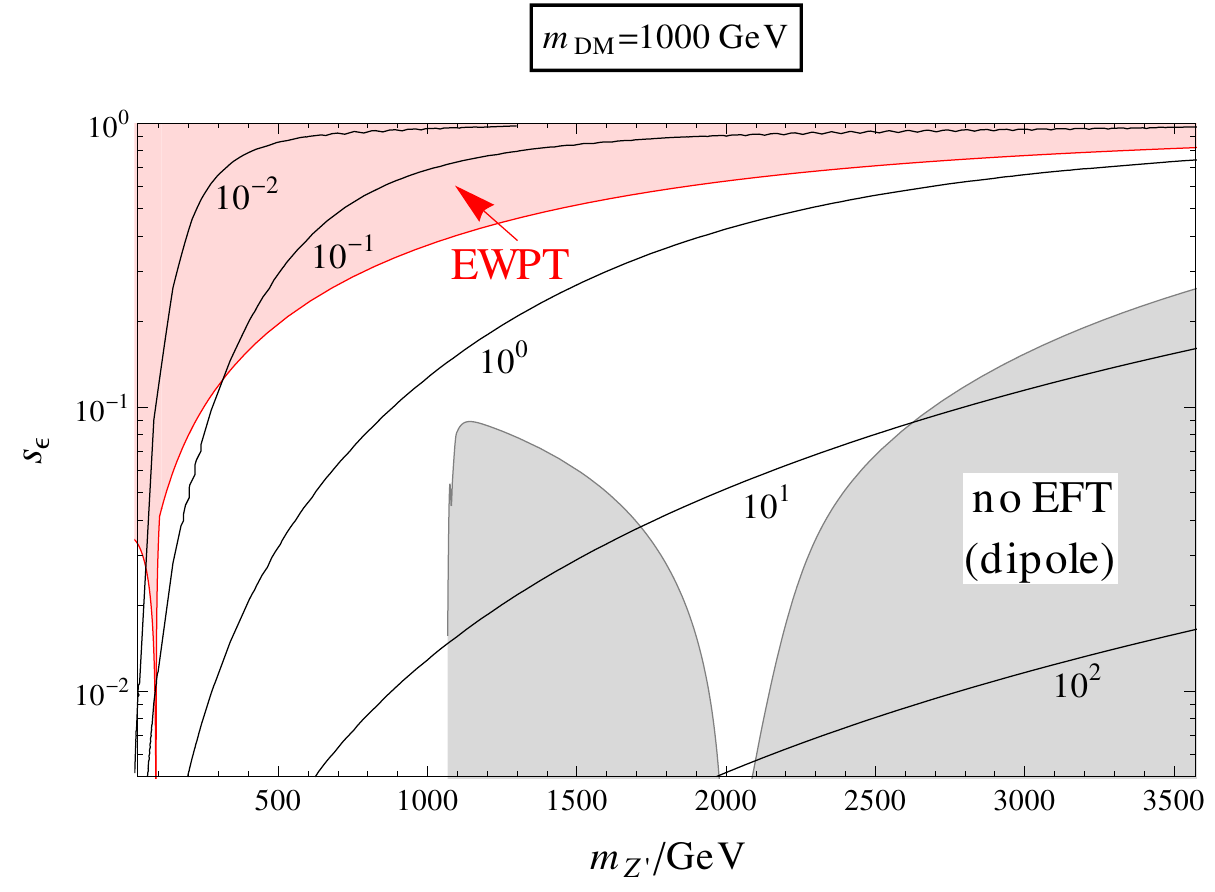}
\caption{Contour plots on the $(\mzp,\se)$ plane for the $90\%$ CL upper limits on $c'$ in the reference model [\eq{refmodel}] from the LUX null result. The dipole and scalar models may be constrained only in the case where effective $U(1)'$ charge operators are generated. In that case the contours are upper limits on $a'$ and $b'$ in \eqs{eqn:effchg}{Phiphi}. For reference, the same EWPT and cosmology constraints as in Fig.~\ref{dipoleplots} are also shown. While EWPT applies to all three models, the ``no EFT'' region is different for different models (see Fig.~\ref{chgrdsplots} for the case of the scalar model).}
    \label{ddplots}
    \end{figure}

    Using the non-relativistic effective operator formalism proposed in~\cite{Fan:2010gt} and further developed in~\cite{Fitzpatrick:2012ix,DelNobile:2013sia}, we calculate the direct detection constraints on $c'$ as a function on the $(\mzp,\se)$ plane (with fixed $\mpsi$) with the help of the Mathematica codes by the authors of~\cite{DelNobile:2013sia}. Among all direct detection experiments, LUX sets the strongest constraints. Thus we show in Fig.~\ref{ddplots} the $90\%$ CL upper limits on $c'$ derived from the LUX null results~\cite{Akerib:2013tjd}.\footnote{Also shown in Fig.~\ref{ddplots} are EWPT and cosmology constraints for the dipole model (same as in Fig.~\ref{dipoleplots}). The latter is irrelevant in the discussion of the reference model here, but will be convenient later when we reinterpret the contours as limits on the dipole model.} As $\se$ increases, the coupling to quarks (nuclei) increases, necessitating a smaller $c'$. The coupling $c'$ constrained as such tends to under-annihilate the dark matter in the early universe for most parts of the parameter space, as suggested by the results in~\cite{Mambrini:2011dw,Kearney:2013xwa}.

    In our models, the tension between direct detection and cosmology is relieved, thanks to the derivatives present in the dipole and charge radius operators. These derivatives turn into the momentum transfer $q$ in dark matter -- nucleon scattering, suppressing the direct detection signal. In the early universe, the derivatives pick up powers of the dark matter mass, and are thus much less suppressed.\footnote{The Rayleigh operator contributes only at loop level and cannot be probed in direct detection, either.}

Effectively, one can think of our models as having ``$c'$-like couplings'' that equal $\frac{\mdm}{\Lambda}$ or $\frac{\mdm^2}{\Lambda^2}$ additionally suppressed by factors of $\frac{-q^2}{\mdm^2}\sim10^{-6}$, which are far below the upper limits in Fig. \ref{ddplots}. Even future direct detection experiments will not be able to probe such small couplings, because the limits on $\epsilon$ in Fig. \ref{ddplots} can at most be pushed down by two orders of magnitude without running afoul of the neutrino background~\cite{Cushman:2013zza}.

However, direct detection may still be interesting for our models, if we consider the possibility that the UV completion of these models may generate an effective $U(1)'$ charge for the dark matter resembling \eq{refmodel}. This can arise from higher dimensional operators involving the hidden Higgs field $\Phi$ responsible for the spontaneous breaking of $U(1)'$. In particular, for the dipole model, consider the operator

   \beq{Phichi}
    i\frac{a}{\Lambda^2}(\Phi^*D_{\mu}\Phi)\bar\chi\gamma^{\mu}\chi +\text{h.c.},
   \eeq
   where  $a$ is a dimensionless coefficient. This operator gives rise to a term
   \begin{equation}
    -\frac{2ag_{Z'}\langle\Phi\rangle^2}{\Lambda^2}\bar\chi\gamma^{\mu}\chi\hat Z'_{\mu} \equiv -a'\bar\chi\gamma^{\mu}\chi\hat Z'_{\mu},\label{eqn:effchg}
   \end{equation}
   where $g_{Z'}$ is the gauge coupling of $U(1)'$. This is just the interaction in \eq{refmodel}, with $c'$ replaced by $a'$, the size of which dpends on the UV theory.
   Thus, the contours in Fig.~\ref{ddplots} can be interpreted as upper limits on the coefficient $a'$ of the effective $U(1)'$ charge operator.

Note that if $a'$ is small (see footnote \ref{footnoteDC}), the operator (\ref{eqn:effchg}) does not significantly change the cosmology, since in the non-relativistic ($s$-wave) limit of $s$-channel $\chi\bar\chi$ annihilation,
   \begin{equation}
   -a'\bar\chi\gamma^{\mu}\chi\hat Z'_{\mu} \simeq -\frac{a'}{4\mchi}\bar\chi\sigma^{\mu\nu}\chi\hat Z'_{\mu\nu}.
   \end{equation}
   The size of this operator is small compared with $\frac{1}{\Lambda}\bar\chi\sigma^{\mu\nu}\chi\hat Z'_{\mu\nu}$, as long as $\Lambda$ is not too much larger than $\mchi$. It has no effect on the ``no EFT'' regions in Fig.~\ref{dipoleplots}.

   Similarly, the scalar model may allow the following effective $U(1)'$ charge operator:
   \beq{Phiphi}
    i\frac{b}{\Lambda^2}(\Phi^*D_{\mu}\Phi)\phi^*\partial^{\mu}\phi +\text{h.c.} = -b'(\phi^*\partial^{\mu}\phi-\phi\partial^{\mu}\phi^*)\hat Z'_{\mu} + \cdots.
   \eeq
   As worked out in the Appendix, at leading order this corresponds to the same non-relativistic effective operator as $-b'\bar\psi\gamma^{\mu}\psi\hat Z'_{\mu}$ in direct detection. Thus the contours in Fig.~\ref{ddplots} may also be interpreted as upper limits on $b'$ for the scalar model.

 \section{The possibility of a light dark matter window}\label{sec:light}

Our focus has been on WIMP dark matter with masses of order $\mathcal{O}(100\text{ GeV})$.  Indeed, for $s$-wave annihilation, data from the CMB and Fermi Large Area Telescope (Fermi-LAT)~\cite{Lopez-Honorez:2013cua,Madhavacheril:2013cna,Ackermann:2011wa} restrict $\mdm\lesssim\mathcal{O}(10\text{ GeV})$, excluding the light dark matter regime.

As we will now discuss, the scalar model considered in this paper at least partially evades such constraints if $\mzp>\mphi$, thanks to the $p$-wave suppression of the only kinetically allowed $2\to 2$ annihilation channel $\phi\bar\phi\to f\bar f$. Here, we will focus on the regime $\mzp>2\mdm$. One motivation is the interesting possibility of tests via ``dark matter beam experiments.'' For example, the dark matter particle may be detected as decay products of $Z^{\prime}$ particles produced in $\pi^0$, $\eta$ decays at MiniBooNE experiments proposed in~\cite{Dharmapalan:2012xp}. While we do not attempt a detailed discussion of parameter space, we demonstrate the existence of a light window for this mass hierarchy, from the perspective of both cosmological and indirect detection considerations. At the masses we have in mind, $\mdm\sim \mathcal{O}(100$ MeV), $g-2$ considerations force $\epsilon < 10^{-3}$; see, e.g.~\cite{Pospelov:2008zw,Essig:2013lka}. If we take $\mdm \sim \mzp$, at the boundary of the validity of the EFT, the only mass scale in the problem is $\mdm$, and the $\phi\bar\phi\to f\bar f$ cross section scales as $\langle \sigma v \rangle \sim (\epsilon/\mdm)^2$.   A naive scaling of the results from Fig.~\ref{chgrdsplots} indicates that $\epsilon ~\gsim 10^{-4}$ should give the correct dark matter abundance.  This implies a window is consistent with the $g-2$ bounds that could usefully be probed by the dark matter beam experiments.

We next discuss the lack of indirect detection constraints for this window.  First, note that the leading subdominant $s$-wave annihilation channel is $\phi \bar{\phi} \to Z^{\prime \, \ast} Z^{\prime \, \ast} \to 4f$. The cross section for this process is suppressed by an additional factor of $10^{-4} \, \epsilon^2 \lesssim 10^{-10}$ relative to the process that determines the relic density (where $10^{-4}$ comes from the phase space), and is easily unconstrained by current data.  It turns out that the $p$-wave annihilation to $f \bar{f}$ is also unconstrained.  Annihilations of dark matter that is virialized in our galaxy are suppressed by $v^{2} \sim 10^{-6}$, and are unconstrained by X-ray and gamma ray data~\cite{Essig:2013goa}. Whether the CMB provides a constraint rests on the velocity of the dark matter at recombination.  The dark matter cools faster once it goes out of kinetic equilibrium, which happens no later than $T\sim$ MeV (after that, the dark matter only scatters off neutrinos, which is suppressed by $q^{4}/m_Z^4$, with $q$ the momentum transfer). By the time of recombination, $v^2\lesssim 10^{-14}$~\cite{Essig:2013goa}, and CMB bounds are easily evaded.

Finally we comment briefly on the case $\mdm<\mzp<2\mdm$.  Here, there may be $s$-wave annihilations to three body final states $\phi \bar{\phi} \to Z^{\prime} f \bar{f}$.  So, we expect less sensitivity to late time annihilations than the dipole model (which has $s$-wave annihilation to two-body final states), but more than in the above case.   Detailed study of this window is left to further work, but it may be possible that a signal could be imprinted in the CMB from these subdominant annihilations.

\section{Conclusions}\label{sec:concl}

We have considered models of dark matter with a $U(1)'$ that kinetically mixes with the SM $U(1)_Y$. In the models considered here, the dark matter interacts only with the $U(1)'$ via higher dimensional operators.  This setup allows a thermal cosmology while avoiding increasingly stringent direct detection constraints.

We find that it is possible to realize a thermal history consistently with the effective theory, whilst simultaneously satisfying constraints from electroweak precision bounds and collider bounds.  However, this is not true for all combinations of $(\mdm, \mzp)$.  Indeed, in the scalar case, the $p$-wave suppression of the process $\phi\bar\phi\to f\bar f$ requires that (1) either the $Z'$ be light enough that the channel $\phi\bar\phi\to Z'Z'$ is open or (2) the dark matter annihilation must be approximately on resonance, $2 \mdm\simeq \mzp$, to partially counteract the $p$-wave suppression.  In either case, it shows the importance of not integrating out the $Z'$ when considering the early universe cosmology.  Even for the dipole case, these two regions make up an important part of the remaining parameter space.

For both models of fermionic and scalar dark matter, LHC searches for the $Z'$ via its decays to leptons represent a powerful probe.  For cases where the $Z'$ can also decay to the dark sector, this branching ratio may be suppressed.  Nevertheless, this is likely the best way to discover such models.  Generally, these models motivate the search for $Z'$'s at the LHC with small production cross sections.

Direct detection is irrelevant for the operators used in this paper to provide a thermal history, due to the derivative interactions. However, it is possible that these models might have observable direct detection signals from other operators that are subdominant for the cosmological history.  Depending on the details of the UV completion, a direct detection signal may be seen at any time.

Finally, we discussed the possibility of a sub-GeV window for the scalar dark matter consistent with CMB constraints, and the low energy experiments sensitive to such a window.  A detailed examination of this window is left for future work.

\appendix*
\section{Some Direct Detection Details}
    In this appendix we discuss the calculation of the direct detection signal both for the reference model and for the dipole and scalar models. For simplicity, we focus on the case $\kappa_{\text{C}}=1$, $\kappa_{\text{R}}=0$ for the scalar model. Our notations are in accord with~\cite{DelNobile:2013sia}.

    The non-relativistic effective operator formalism in~\cite{Fan:2010gt,Fitzpatrick:2012ix,DelNobile:2013sia} is based on the fact that the dark matter - nucleon interactions in direct detection experiments can be described by a set of 12 non-relativistic effective operators $\mathcal{O}_i^{\text{NR}}$. The lagrangian of any dark matter model can be reduced to the sum of these operators in the non-relativistic limit:
    \beq{LNR}
    \Lag_{\text{eff}}^{\text{NR}}=\sum\limits_{i=1}^{12}\sum\limits_{N=p,n}\mathfrak{c}_i^N\mathcal{O}_i^{\text{NR}},
    \eeq
    where the proton and the neutron may contribute differently. Then, the theoretical prediction for the number of signal events $N^{\text{th}}$ is determined by the coefficients $\mathfrak{c}_i^N$ as follows:
    \begin{equation}
    \label{Nth}
    \begin{split}
    N^{\text{th}} &= \frac{\rho_{\text{DM}}}{m_{\text{DM}}}\frac{1}{32\pi}\frac{1}{m_{\text{DM}}^2m_N^2}\sum_{i,j=1}^{12} \sum_{N,N'=p,n} \mathfrak{c}_i^N(\{\lambda\},m_{\text{DM}})\mathfrak{c}_j^{N'} (\{\lambda\},m_{\text{DM}})\tilde{\mathcal{F}}_{i,j}^{(N,N')}(m_{\text{DM}}),
    \end{split}
    \end{equation}
    where $\rho_{\text{DM}}\simeq 0.3\text{ GeV/cm}^3$, and $\{\lambda\}$ represents all parameters of the dark matter model. $\tilde{\mathcal{F}}_{i,j}^{(N,N')}$ are the integrated nuclear form factors convolved with all the experimental effects, which characterize the target's response to dark matter. They depend on the experimental condition, but not on the dark matter model.  The null results of direct detection experiments thus set limits on $\{\lambda\}$.

    To derive $\Lag_{\text{eff}}^{\text{NR}}$ (and hence $\mathfrak{c}_i^N$) from the relativistic model Lagrangian $\Lag_{\text{DM}}$ given in \eqss{refmodel}{eq:dipole}{eq:scalar} is a two-step process. First, we integrate out the heavy gauge bosons $Z'$ and $Z$ being exchanged in the scattering to obtain a relativistic effective Lagrangian of current interactions:
    \be
    \label{Leff}
    \Lag_{\text{eff}}= \frac{\se\cw}{\ce^2(1+\td\te\sw)}\frac{1}{\mzp^2}J_{Z'\mu}J_{\text{EM}}^{\mu} \simeq \frac{\se\cw}{\ce^2\mzp^2}J_{Z'\mu}J_{\text{EM}}^{\mu},
    \ee
    where
    \be
    \label{jzp}
     J_{Z'}^{\mu} =
    \begin{cases}
    \frac{c'}{2\mdm} \bar\psi (P^{\mu} -  i\sigma^{\mu\nu} q_{\nu}) \psi  \hspace{0.5in} &\text{(reference model)}\\
    -\frac{2}{\Lambda}\bar\chi i\sigma^{\mu\nu} q_{\nu}\chi &\text{(dipole model)}\\
       \frac{-q^2}{2\Lambda^2}P^{\mu}\phi^*\phi &\text{(scalar model -- charge radius only),}
    \end{cases}
    \ee
    \be
    \label{jem}
    J_{\text{EM}}^{\mu} = \frac{e}{2m_N} \sum_{N=p,n} \bar N \left(Q_N K^{\mu} + \frac{g_N}{2} i\sigma^{\mu\nu} q_{\nu}\right) N.
    \ee
    \eq{jzp} can be derived easily from \eqss{refmodel}{eq:dipole}{eq:scalar}. $Q_p=1$, $Q_n=0$, $g_p=5.59$, $g_n=-3.83$ in \eq{jem} are the electric charges and magnetic $g$-factors of the nucleons. The convention for the momenta is: $p^{\mu}$ and $p'^{\mu}$ ($k^{\mu}$ and $k'^{\mu}$) are the incoming and outgoing momenta of the dark matter (nucleon), $P^{\mu}=p^{\mu}+p'^{\mu}$, $K^{\mu}=k^{\mu}+k'^{\mu}$,  $q^{\mu}=p'^{\mu}-p^{\mu}=k^{\mu}-k'^{\mu}$. Note that $ J_{Z'}^{\mu}$ couples to $J_{\text{EM}}^{\mu}$, but not to $ J_{Z}^{\mu}$, in the low $q^2$ limit. This is most easily seen in the $(\hat Z, \hat A, \hat Z')$ basis, where the two contributing diagrams involve $\hat Z'$-$\hat Z$ and $\hat Z'$-$\hat A$ kinetic mixing, respectively, in the $t$-channel propagator. The vertices for kinetic mixing are comparable for the two diagrams, but the first diagram is suppressed by the $\hat Z$ propagator compared with the second, which explains the absence of $ J_{Z}^{\mu}$ in \eq{Leff}.

    Now with $\Lag_{\text{eff}}$ at hand, we proceed to the second step: taking the non-relativistic limit of $\Lag_{\text{eff}}$. Following the methods of~\cite{Fitzpatrick:2012ix,DelNobile:2013sia}, we obtain $\Lag_{\text{eff}}^{\text{NR}}$ as in \eq{LNR}, with the following nonvanishing coefficients:
    \begin{itemize}
    \item Reference model: $\mathfrak{c}_1^p = c' \mathscr{C}\left[1-\left(\frac{1}{4\mdm^2}+\frac{g_p}{8m_N^2}\right)q^2\right]$, $\mathfrak{c}_1^n = c' \mathscr{C}\left(-\frac{g_n}{8m_N^2}q^2\right)$, $\mathfrak{c}_3^{p,n} =  c' \mathscr{C}\left(-\frac{g_{p,n}}{2m_N}\right)$, $\mathfrak{c}_4^{p,n} =  c' \mathscr{C}\left(-\frac{g_{p,n}}{2m_N\mdm}q^2\right)$, $\mathfrak{c}_5^p =  c' \mathscr{C}\left(-\frac{1}{\mdm}\right)$, $\mathfrak{c}_6^{p,n} =  c' \mathscr{C}\left(\frac{g_{p,n}}{2m_N\mdm}\right);$
    \item Dipole model: $\mathfrak{c}_1^p = \left(\frac{4\mdm}{\Lambda}\right) \mathscr{C}\left(-\frac{1}{4\mdm^2}q^2\right)$, $\mathfrak{c}_4^{p,n} =  \left(\frac{4\mdm}{\Lambda}\right) \mathscr{C}\left(-\frac{g_{p,n}}{2m_N\mdm}q^2\right)$, $\mathfrak{c}_5^p =  \left(\frac{4\mdm}{\Lambda}\right) \mathscr{C}\left(-\frac{1}{\mdm}\right)$, $\mathfrak{c}_6^{p,n} =  \left(\frac{4\mdm}{\Lambda}\right)\mathscr{C}\left(\frac{g_{p,n}}{2m_N\mdm}\right);$
    \item Scalar model (charge radius only): $\mathfrak{c}_1^p = \left(\frac{-q^2}{2\Lambda^2}\right) \mathscr{C}\left(1-\frac{g_p}{8m_N^2}q^2\right)$, $\mathfrak{c}_1^n = \left(\frac{-q^2}{2\Lambda^2}\right) \mathscr{C}\left(-\frac{g_n}{8m_N^2}q^2\right)$, $\mathfrak{c}_3^{p,n} =  \left(\frac{-q^2}{2\Lambda^2}\right) \mathscr{C}\left(-\frac{g_{p,n}}{2m_N}\right).$
    \end{itemize}
    Here $\mathscr{C} = \frac{4e\se\cw m_N\mdm}{\ce^2\mzp^2}$.

    As calculated in~\cite{DelNobile:2013sia}, $\tilde{\mathcal{F}}_{1,1}^{(N,N')}$ are at least 3 orders of magnitude larger than all other $\tilde{\mathcal{F}}_{i,j}^{(N,N')}$'s (dimensionful $\tilde{\mathcal{F}}$'s are in units of GeV). Further, $|q^2|\ll m_N, m_{\text{DM}}$. Thus, for the reference model, to a good approximation, only $\mathfrak{c}_1^p \simeq c' \mathscr{C}$ needs to be kept in \eq{Nth}, and it is $q^2$ independent.\footnote{This is  the standard spin-independent (SI) interaction.} This simplification allows us to easily calculate the expected direct detection signal $N^{\text{th}}$ for the reference model using the Mathematica codes by the authors of~\cite{DelNobile:2013sia}.\footnote{These codes can only deal with models with $q^2$ independent $\mathfrak{c}_i^N$, and thus are not directly useful for calculating our models. This is why we take an indirect approach, namely to study the reference model first and then work out the suppression in our models compared with the reference model.} The resulting upper limits on $c'$ due to the LUX null results are shown in Fig.~\ref{ddplots}.

    We can directly translate these results to the dipole model, by noting that the equivalent of $c'$ here is $\left(\frac{4\mchi}{\Lambda}\right)$ suppressed by the following factors:

    \begin{equation}
    \begin{split}
    & \text{For $\mathfrak{c}_1^p$: }\,\,\,\,\frac{-q^2}{4\mchi^2} \sim 10^{-6},\\
    & \text{For $\mathfrak{c}_4^{p,n}$: }\frac{-g_{p,n}q^2}{2m_N\mchi}\sqrt{\mathcal{Y}_{4,4}^{N,N'}} \lesssim 10^{-6},\\
    & \text{For $\mathfrak{c}_5^p$: }\,\,\,\,\frac{\text{GeV}}{\mchi}\sqrt{\mathcal{Y}_{5,5}^{p,p}} \lesssim 10^{-6},\\
    & \text{For $\mathfrak{c}_6^{p,}$: }\frac{g_{p,n}(\text{GeV})^2}{2m_N\mchi}\sqrt{\mathcal{Y}_{6,6}^{N,N'}} \lesssim 10^{-7}.
    \end{split}
    \end{equation}
    where $\mathcal{Y}_{i,j}^{N,N'}=\tilde{\mathcal{F}}_{i,j}^{N,N'}/\tilde{\mathcal{F}}_{1,1}^{p,p}$ are plotted in~\cite{DelNobile:2013sia}. These suppressions explain the absence of direct detection signals, as explained in Section~\ref{sec:dd}.

    Similar conclusions can be reached for the scalar model. Here, as in the reference model, only $\mathfrak{c}_1^p \simeq \left(\frac{-q^2}{2\Lambda^2}\right) \mathscr{C} = \left(\frac{-q^2}{2\mphi^2}\right)\left(\frac{\mphi^2}{\Lambda^2}\right) \mathscr{C}$ needs to be kept, and the suppression factor is seen to be $\frac{-q^2}{2\mphi^2}\sim10^{-6}$. We also note that the effective $U(1)'$ charge operator in \eq{Phiphi} gives rise to $J_{Z'}^{\mu}=b'P^{\mu}\phi^*\phi$. Thus the calculation is exactly the same as the charge radius operator, with $\frac{-q^2}{2\Lambda^2}$ replaced by $b'$, leading to $\mathfrak{c}_1^p \simeq b'\mathscr{C}$ (with other $\mathfrak{c}_i^N$ negligible) at leading order. This resembles the nonrelativistic limit of $-b'\bar\psi\gamma^{\mu}\psi\hat Z'_{\mu}$.

\acknowledgments
We acknowledge useful conversations with B.~Batell, L.~Fitzpatrick, L.~Hall, J.~Kearney, M.~Papucci, J.~Wacker, and J.~Wells.  The work of AP supported by DoE grant DE-SC0007859 and the work of ZZ is supported by CAREER grant NSF-PHYS 0743315.

\bibliography{DM}
\end{document}